\documentclass[10pt,journal,twoside]{IEEEtran}

\usepackage{amsmath,amssymb,amsthm}
\usepackage{mathrsfs}
\usepackage{latexsym}
\usepackage{bm}
\usepackage{cases}
\usepackage{balance}
\usepackage{graphicx}
\usepackage{subfigure}
\usepackage{float}

\usepackage{color}
\usepackage{accents}
\usepackage{enumerate}
\usepackage{cite}
\usepackage{CJK}

\usepackage{longtable}
\usepackage{diagbox}   
\usepackage{multirow}
\usepackage{arydshln}  
\usepackage{booktabs}
\usepackage{tikz}
\usetikzlibrary{calc,positioning,fit,matrix,arrows.meta,decorations.pathreplacing}
\tikzset{
  box/.style = {draw, rectangle, minimum width=18mm, minimum height=7mm, align=center},
  circ/.style = {draw, circle, minimum size=7mm, inner sep=1pt, align=center},
  arr/.style = {-{Stealth[length=3mm]}, line width=0.6pt}
}

\usepackage{algorithm}
\usepackage{algpseudocode}
\usepackage{xpatch}

\makeatletter
\def\widebar{\accentset{{\cc@style\underline{\mskip10mu}}}}
\def\Widebar{\accentset{{\cc@style\underline{\mskip8mu}}}}
\makeatother

\allowdisplaybreaks

\theoremstyle{plain}

\newtheorem{lemma}{Lemma}

\theoremstyle{definition}
\theoremstyle{definition}

\algrenewcommand\algorithmicrequire{\textbf{Input:}}
\algrenewcommand\algorithmicensure{\textbf{Output:}}
\def\BibTeX{{\rm B\kern-.05em{\sc i\kern-.025em b}\kern-.08em
    T\kern-.1667em\lower.7ex\hbox{E}\kern-.125emX}}

\begin{document}
\title{\huge Affine Frequency Division Multiplexing With Pre-Chirp Spreading Index Modulation: A CSI-Driven Secure Transmission Framework}
\author{Yiwei~Tao,~\IEEEmembership{Graduate Student Member,~IEEE},
        Miaowen~Wen,~\IEEEmembership{Senior Member,~IEEE}, Yao~Ge,~\IEEEmembership{Senior Member,~IEEE}, 
        Merouane~Debbah,~\IEEEmembership{Fellow,~IEEE},
        and Erdal~Panayirci,~\IEEEmembership{Life Fellow,~IEEE}\vspace{-1mm}
\thanks{The work of M. Wen was supported in part by the National Key R\&D Program of China under Grant 2023YFB2904500 and in part by the Fundamental Research Funds for the Central Universities under Grant 2026ZYGXZR031. The work of Yao Ge is supported by A*STAR under the RIE2025 Industry Alignment Fund--Industry Collaboration Projects (IAF-ICP) Funding Initiative (Award: I2501E0045), as well as cash and in-kind contribution from the industry partner(s). \emph{(Corresponding author: Miaowen~Wen.)}}
\thanks{Yiwei Tao and Miaowen Wen are with the Guangdong Provincial Key Laboratory of Short-Range Wireless Detection and Communication, School of Electronic and Information Engineering, South China University of Technology, Guangzhou 510640, China (e-mail: eeyiweitao@mail.scut.edu.cn; eemwwen@scut.edu.cn).}%
\thanks{Yao Ge is with the AUMOVIO-NTU Corporate Lab, Nanyang Technological University, Singapore 639798 (e-mail: yao.ge@ntu.edu.sg).}%
\thanks{Merouane Debbah is with the Research Institute for Digital Future, Khalifa University, 127788 Abu Dhabi, UAE (e-mail: merouane.debbah@ku.ac.ae).}%
\thanks{Erdal Panayirci is with the Department of Electrical and Electronics Engineering, Kadir Has University, Istanbul 34230, T\"{u}rkiye (e-mail: eepanay@khas.edu.tr).}%
}
\markboth{}%
{\MakeLowercase{\textit{}}}


\maketitle

\vspace{-0.1cm}
\begin{abstract}
The flexible configurability of pre-chirp parameters in affine frequency division multiplexing (AFDM) provides new degrees of freedom for designing covert communication mechanisms. Leveraging this property, this paper proposes a covert AFDM pre-chirp spreading index modulation (IM) framework referred to as {\em{AFDM-PCS-IM}}, tailored for high-mobility scenarios.
Specifically, an orthogonal pre-chirp parameter codebook is constructed by exploiting the tunability of the pre-chirp parameters. To enhance covertness, the legitimate channel state information (CSI) is utilized to generate controlled perturbations that encrypt the codebook, thereby tightly coupling the modulation structure with instantaneous channel characteristics. 
The encrypted pre-chirp codebook is then employed as an IM dimension for information transmission, where the index bits are conveyed through the selection of pre-chirp parameter sequences. Since eavesdroppers cannot obtain consistent CSI, they are unable to reconstruct the encrypted codebook and thus cannot reliably recover any useful information.
Moreover, to address channel estimation errors inherent in practical communication environments, a noise-aware CSI quantizer is further introduced to improve the robustness under imperfect CSI conditions. Finally, a low-complexity two-layer message passing detector with Walsh code constraints (TLMP-WC) is developed to enable efficient large-scale data detection while simultaneously enhancing the detection reliability.
Simulation results demonstrate that the proposed AFDM-PCS-IM framework with the TLMP-WC detector achieves superior covertness, improved bit error rate (BER) performance, and reduced computational complexity compared with the state-of-the-art schemes.
\end{abstract}

\section{Introduction}
With the rapid advancement of wireless technologies and the increasingly dense deployment of mobile devices, wireless services have expanded into high-mobility scenarios such as high-speed railways, unmanned aerial vehicles (UAVs), and low-Earth-orbit (LEO) satellites~\cite{10794592,11003079,11556275}. In such environments, not only reliable transmission but also covert communication becomes significantly more challenging, as rapidly varying channels undermine key consistency and channel-dependent secrecy mechanisms. These challenges highlight the need for waveforms that maintain robustness under high mobility while supporting the integration of physical-layer security (PLS) mechanisms~\cite{10384384}.

PLS mechanisms have gained increasing attention as they leverage legitimately obtained channel state information (CSI) to couple waveform and modulation design with channel characteristics, enabling inherent confidentiality without relying on computational cryptography~\cite{10131983}. To fully exploit PLS in highly dynamic environments, however, the underlying waveform must maintain structural stability under rapid channel variations.
So far, affine frequency division multiplexing (AFDM) has recently emerged as a promising candidate, owing to its inherent robustness to severe delay and Doppler shifts and its suitability for high-mobility communications~\cite{10087310}. This robustness stems from the flexible chirp structure of AFDM, which enables fine delay-Doppler resolution and allows the waveform to attain full time-frequency diversity over linear time-varying (LTV) channels. As a result, AFDM exhibits superior resilience to doubly selective fading compared with the traditional orthogonal frequency division multiplexing (OFDM) and other LTV-oriented waveforms, such as orthogonal chirp division multiplexing (OCDM)~\cite{7523229}, orthogonal delay-Doppler division multiplexing (ODDM)~\cite{9829188}, and orthogonal time frequency space~(OTFS) \cite{7925924,WeiTransmitterandReceiver2021,LiPerformanceAnalysis2021}.


\subsection{Related Work}
For these advantages, AFDM has attracted increasing research attention in recent years~\cite{10769778,11204797}.
Early studies established the fundamental AFDM modulation framework and derived its input-output relationships in both the discrete affine Fourier transform (DAFT) domain and time domain, revealing that AFDM preserves the inherent sparsity of doubly selective channels and enables full delay-Doppler diversity~\cite{9473655}.
Building upon this framework, extensive efforts have been devoted to receiver design and performance enhancement, including iterative soft-MMSE detection~\cite{11214369}, joint sparse graph receiver design~\cite{11150613}, and cross-domain expectation propagation detectors~\cite{11303728,11185309}.
These low-complexity equalization and detection schemes, which explicitly leverage the structured sparsity in the DAFT domain, effectively reduce computational burden while maintaining near-optimal performance~\cite{9349154,9354639,11027148}.
In parallel, channel estimation strategies tailored to AFDM were developed to ensure reliable data detection in high-mobility environments.
For example, various schemes such as embedded pilot designs and data-pilot superposition transmission were proposed to balance spectral efficiency and estimation reliability~\cite{10711268,11315874}.
Recent work has also investigated MIMO-AFDM under practical transceiver
hardware impairments, demonstrating its robustness to both multiplicative
and additive impairments and its clear BER advantages over MIMO-OFDM
under nonideal transceiver conditions~\cite{Sui2026Hardware}.
These efforts further demonstrated the suitability of AFDM for channels with severe Doppler dispersion, highlighting the effectiveness of its structured delay-Doppler representation\cite{10439996}.
This structural characteristic also establishes a natural bridge to sensing functionality, as the delay-Doppler domain directly corresponds to the range-velocity parameter space in radar systems, thereby enabling integrated sensing and communications (ISAC) within a unified signal framework~\cite{11173628,11271365,11164334}.
More broadly, recent work has placed AFDM within a unified chirp-signaling
framework that connects sequences, multicarrier waveforms, high-mobility
communications, sensing, and ISAC, thereby highlighting its wider
relevance to next-generation wireless systems~\cite{Sui2025Chirp}.

Beyond point-to-point transmission, AFDM has also been extended at both the waveform and system levels.
By analyzing the impact of DAFT per-chirp parameters on the signal envelope characteristics, prior work demonstrated that proper pre-chirp parameter configuration can significantly reduce the peak-to-average power ratio (PAPR) of AFDM~\cite{10901277}.
This observation motivated further studies on per-chirp parameter optimization for enhanced waveform efficiency, ranging from analytical design methods to data-driven techniques such as deep learning-based optimization for PAPR reduction~\cite{10812762,11299064}.
At the system level, multi-user transmission schemes were designed to support multiple access over doubly selective channels~\cite{10566604,11185309}.
Moreover, index modulation (IM) techniques were incorporated into AFDM to enhance spectral efficiency and transmission flexibility~\cite{tao2023affine,tao_JSAC}.
Following this direction, generalized spatial modulation-aided AFDM was
developed for reliable MIMO communications over doubly selective channels,
together with tailored low-complexity detectors and an analytical BER
upper bound~\cite{11185315}, while IM-assisted AFDM was further extended
to the constellation~\cite{11112817} and per-chirp
parameter~\cite{10975107} domains.
Despite these significant advances in reliability, efficiency, and functionality, existing studies have primarily focused on communication performance metrics such as bit error rate (BER), spectral efficiency (SE), and detection complexity.

\vspace{-0.2cm}
\subsection{Motivation and Contribution}
In contrast, the PLS aspects of AFDM systems remain relatively underexplored. 
Several preliminary efforts have indicated that the flexible configurability of AFDM pre-chirp parameters can be exploited for security enhancement~\cite{11393557}. 
Specifically, a chirp-permutation-based security was investigated in~\cite{11003930}, where the combinatorial search space induced by pre-chirp reordering is utilized for data encryption. A parameter-hopping strategy was proposed in~\cite{11392818} to dynamically vary waveform configurations, thereby enlarging the effective parameter space and increasing waveform entropy. 
The intrinsic robustness of AFDM against brute-force demodulation was analyzed in~\cite{11345476}, showing that its continuous parameter domain results in higher search complexity than OTFS. 
In addition, a channel-reciprocity-based secure AFDM framework was developed in~\cite{11072374}, in which pre-chirp parameters are generated from shared channel features between legitimate transceivers.
However, existing AFDM security methods mainly employ pre-chirp
parameters as permutation, hopping, or directly channel-derived secure
waveform variables, rather than as structured information-bearing
codebooks~\cite{11003930,11392818,11345476,11072374}. In particular, the joint
design of an orthogonal pre-chirp codebook for index modulation, its
CSI-driven encryption, and a corresponding structure-aware
low-complexity detector has not been fully investigated.

Motivated by this insight, we revisit the flexible configurability of AFDM pre-chirp parameters from the modulation structure perspective. 
Instead of treating them merely as secret keys or hopping variables, we exploit them as an additional index dimension to reshape the waveform structure. 
Accordingly, an AFDM pre-chirp spreading index modulation framework is proposed, referred to as {\em AFDM-PCS-IM}. 
We first leverage the pre-chirp parameters to construct an orthogonal pre-chirp parameter codebook. 
Then, legitimate CSI is utilized to generate controlled perturbations to encrypt the pre-chirp parameter codebook, thereby tightly coupling the modulation structure with instantaneous channel characteristics. 
The proposed AFDM-PCS-IM framework significantly enhances the covertness performance while maintaining reliable transmission in high-mobility scenarios. The key contributions of this work are summarized as~follows:
\begin{itemize}

\item
We propose an AFDM-PCS-IM framework by designing an orthogonal codebook construction rule that generates multiple mutually orthogonal sets of pre-chirp parameters. In the proposed structure, information bits are jointly carried by the modulation constellation symbols and the selected pre-chirp parameter patterns, fully exploiting the degrees of freedom in the pre-chirp parameter domain.

\item
We propose an orthogonal pre-chirp parameter codebook and a CSI-driven perturbation sequence generation method. The perturbation sequence is applied to the pre-chirp parameter codebook to encrypt the data, while ensuring that legitimate users can still correctly distinguish the intended signals by using the channel reciprocity. This greatly improves the covertness of the AFDM-PCS-IM scheme and makes it hard for eavesdroppers to obtain useful information without the legitimate CSI.

\item
We propose a noise-aware CSI quantization mechanism to mitigate the impact of channel estimation errors in practical communication environments. 
The proposed mechanism can improve robustness under imperfect CSI conditions, ensuring reliable encrypted codebook reconstruction at the legitimate receiver in high-mobility environments.

\item
We develop a low-complexity two-layer message passing detector with Walsh code constraints (TLMP-WC) to enable efficient large-scale data detection. The proposed TLMP-WC detector enhances detection reliability while maintaining reduced computational complexity compared with existing detectors. Simulation results verify that the proposed AFDM-PCS-IM scheme with TLMP-WC detector achieves superior covertness, reliability, and robustness compared with state-of-the-art methods.

\end{itemize}

The rest of this paper is organized as follows. Section II presents the system model of the proposed AFDM-PCS-IM scheme. Section III describes the pre-chirp parameter codebook design and encryption principles. Section IV introduces a robustness mechanism based on CSI quantization. Section V introduces the principle of the proposed TLMP-WC detector.  Section VI provides simulation results and performance comparisons with existing benchmarks. Finally, Section VII concludes the~paper.

\begin{figure*}[t]
	\center
	\includegraphics[width=6.3in,height=2.3in]{{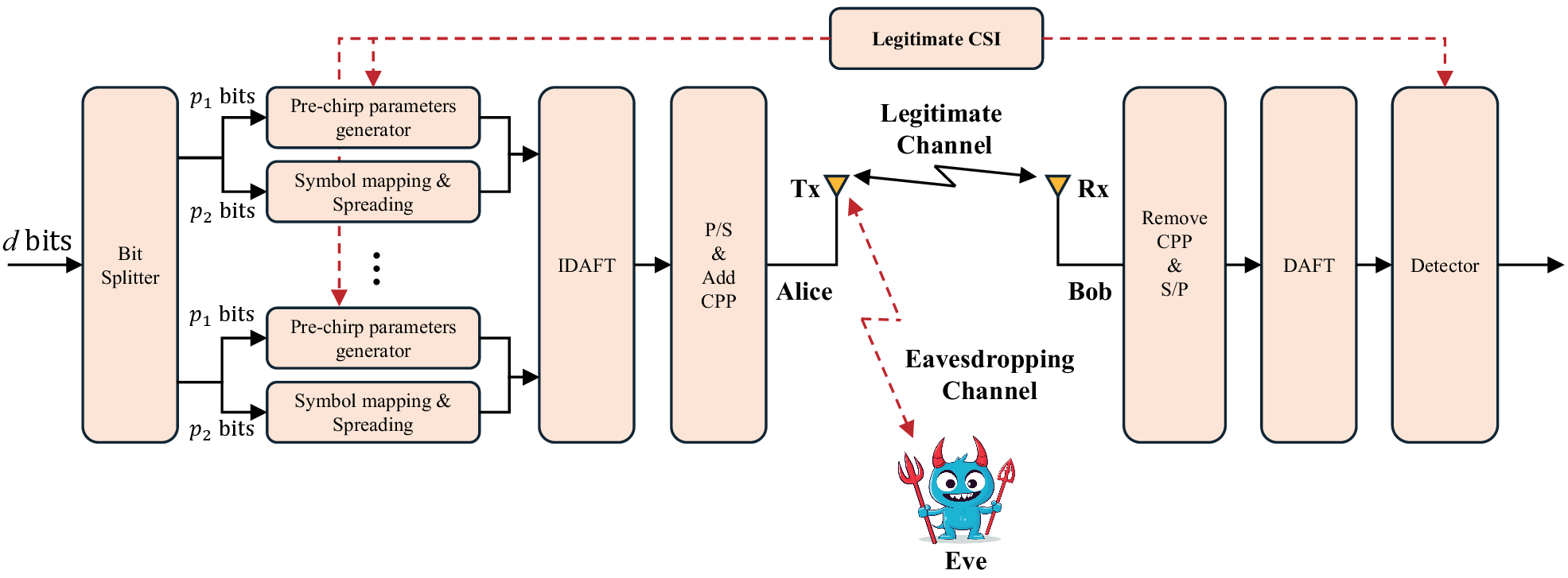}}
	\vspace{-0.3cm}
	\caption{{Transceiver structure of the proposed AFDM-PCS-IM scheme.}}
	\label{system-model}  
	\vspace{-4mm}
\end{figure*}
{\em{Notations:}}
The operators $(\cdot)^{T}$, $(\cdot)^{H}$, and $(\cdot)^{-1}$ denote the transpose, conjugate transpose, and matrix inversion, respectively.
The symbol $\lfloor \cdot \rfloor$ represents the floor function.
The binomial coefficient $C(n,m)$ denotes the number of ways to select $m$ elements from $n$ elements.
The operators $[\cdot]_{N}$ and $\|\cdot\|$ denote modulo-$N$ and the Euclidean norm, respectively.
The notation $\mathrm{diag}(\cdot)$ forms a diagonal matrix from a given vector.
The sets $\mathbb{C}^{M \times N}$ and $\mathbf{I}_{N}$ represent an $M \times N$ complex-valued matrix and an $N \times N$ identity matrix, respectively.
The operator $\arg(\cdot)$ returns the phase of a complex number.
The notation ``$\mathrm{mod}\, 1$'' indicates that the result is taken modulo $1$.
The notation $\|\mathbf{x}\|_0$ denotes the number of non-zero elements in $\mathbf{x}$.

\section{System Model}
As shown in Fig.~\ref{system-model}, we consider a communication model consisting of a legitimate transmitter (Alice), receiver (Bob), and a potential eavesdropper (Eve), in which Alice transmits signals using the AFDM waveform. 
Assume a reciprocity-calibrated time-division duplexing (TDD)
system, in which Alice and Bob locally acquire estimates of the
legitimate CSI through bidirectional pilot signaling~\cite{11072374,Shi2026Pilot}. Specifically,
Alice estimates the reciprocal Alice$\to$Bob LTV channel from the uplink
pilot transmitted by Bob, whereas Bob estimates the forward
Alice$\to$Bob channel from a pilot or preamble transmitted by Alice.
Consequently, neither the instantaneous CSI nor the CSI-driven
perturbation sequence needs to be explicitly exchanged between Alice
and Bob.
In contrast, Eve is located at a different position and cannot reliably acquire the legitimate CSI in high-mobility scenarios.

\subsection{Fundamentals of AFDM}
AFDM is a waveform based on the discrete affine Fourier transform (DAFT), which has two adjustable parameters: the post-chirp parameter $c_1$ and the pre-chirp parameter $c_2$. Its matrix form can be expressed as follows
\begin{align}\label{DAFT}
  {\bf A}={\bm\Lambda}_{c_2}{ {\bf F}_{N}}{\bm \Lambda}_{c_1},
\end{align}
where ${{\bf F}_{N}}$, ${\bm\Lambda}_{c_1}$, and ${\bm\Lambda}_{c_2}\in{\mathbb{C}}^{N\times N}$ are the $N$-point normalized fast Fourier transform (FFT) matrix, post-chirp matrix, and pre-chirp matrix, respectively. The matrix ${{\bm\Lambda}_{c\in\{c_1,c_2\}}}$ is given by
\begin{align}\label{matrix}
{\bf\Lambda}_{{{c}}}={\rm diag}\left(e^{-j2\pi {c}{n}^2}, n=0,\ldots,N-1\right).
\end{align}
The post-chirp parameter $c_1$ is set to ${(2{\alpha}_{\rm max}+1+{k}_{\alpha}})/{2N}$ to ensure full delay-Doppler diversity in the DAFT domain, where ${\alpha}_{\rm max}$ denotes the maximum normalized integer Doppler over the LTV channel, and ${k}_{\alpha}$ is an adjustable parameter that reduces the performance loss caused by fractional Doppler dispersion. The pre-chirp parameter $c_2$ is typically set to an arbitrary irrational number or a value less than $1/{2N}$~\cite{10087310}.

{\bf {Remark 1:}} In conventional AFDM systems, it is commonly assumed that all chirp subcarriers share the same pre-chirp parameter $c_2$, thereby leaving the degrees of freedom offered by the pre-chirp parameters largely underexploited. To overcome this limitation, we allow different pre-chirp parameters to be assigned across chirp subcarriers and design a pre-chirp parameter codebook to fully exploit these degrees of freedom in~Section~\ref{codebook} for performance enhancement.


\subsection{Transmitter of the Proposed AFDM-PCS-IM}
At the transmitter, the $N$ chirp subcarriers of an AFDM-PCS-IM symbol are divided into $G$ groups, each containing $M=N/G$ chirp subcarriers. Likewise, the $d$ information bits are divided into $G$ groups, with each AFDM sub-group carrying $p = d/G$ bits. As each group undergoes similar operations, the $g$-th group is taken as an example for exposition, where $g\in{\{1,2,\ldots,G\}}$.

First, we represent the pre-chirp parameters $c_2$ of the $M$ subcarriers in the $g$-th group as a pre-chirp sequence ${\bf c}_{2,g}=[{c}_{2,g,0},\ldots,{c}_{2,g,m},\ldots,{c}_{2,g,M-1}]$, where ${c}_{2,g,m}$ denotes the pre-chirp parameter of the $m$-th subcarrier in the $g$-th group.
By exploiting the flexibility of the pre-chirp parameters and legitimate CSI, a pre-chirp parameter codebook can be constructed, i.e.,  ${\mathcal C}_{2,g}=[{\bf c}_{2,g}^{1};\ldots;{\bf c}_{2,g}^{k};\ldots;{\bf c}_{2,g}^{K}]$, and subsequently employed to establish the IM rule.
The detailed description of the pre-chirp parameter codebook design will be discussed in Section~\ref{codebook}.
The pre-chirp parameter codebook is composed by $K$ pre-chirp sequences, where ${\bf c}_{2,g}^{k}$ represents the $k$-th pre-chirp~sequence.
Next, the $p$ bits are divided into $p_1$ index bits and $p_2$ modulated bits. The $p_1$ bits are mapped to the selection of the pre-chirp sequence ${\bf c}_{2,g}\in {\mathcal C}_{2,g}$. For example, when $M=K=4$, the per-chirp sequence embedded in the $g$-th group chirp subcarriers is given in Table~\ref{table1}, which has four possible choices. Therefore, the number of index bits $p_1={\rm log}_2\left\lfloor C(K,1)\right\rfloor$.
The remaining $p_2={\rm log}_2(\mathcal Q)$ modulated bits are mapped to a $\mathcal Q$-ary constellation symbol $x_g$ (e.g., $\mathcal Q$-ary PSK).
The symbol $x_g$ is then spread over the $M$ chirp subcarriers in the $g$-th group using an all-one spreading vector ${\bf 1}_{M}\in{\mathbb{R}^{1\times M}}$, so that each subcarrier in the group transmits the same symbol~$x_g$.

By performing the same operations for all $G$ groups, the DAFT-domain transmitted symbol vector ${\bf x}\in \mathbb{C}^{N \times 1}$ and the corresponding pre-chirp parameter sequence ${\bf c}_{2}^{{\rm A}}$ in Alice can be constructed as follows
\begin{align}\label{tran-x}
  {\bf x}=[{x_1}{\bf 1}_{M},\ldots,{x_g}{\bf 1}_{M},\ldots,{x_G}{\bf 1}_{M}]^{T},
\end{align}
\begin{align}\label{c2}
  {\bf c}_{2}^{{\rm A}}=[{\bf c}_{2,1}^T,\ldots,{\bf c}_{2,g}^T,\ldots,{\bf c}_{2,G}^T]^T.
\end{align}
Subsequently, we perform the inverse DAFT (IDAFT) to convert the transmitted vector symbol $\bf x$ into the time-domain transmitted signal ${\bf s}\in \mathbb{C}^{N \times 1}$, i.e.,
\begin{align}\label{transmitted}
  {\bf s}={{\bm\Lambda}_{c_1}^H{ {\bf F}_{N}^H}{\bm \Lambda}_{{c_{2}^{{\rm A}}}}^H{\bf x}},
\end{align}
where ${\bm \Lambda}_{{c_{2}^{{\rm A}}}}={\rm diag}\left(e^{-j2\pi {{c_{2}^{{\rm A}}}[n]}{n}^2}, n=0,\ldots,N-1\right)$,
and ${{c_{2}^{{\rm A}}}[n]}$ is the $n$-th element in ${\bf c}_{2}^{{\rm A}}$.
Finally, to avoid inter-block interference, we add a chirp periodic prefix (CPP)~\cite{10087310} to the time-domain transmitted signal and then send it through the LTV channel to the Bob.

\begin{table}[t]
\caption{{An Example of the Proposed IM Scheme With $M=K=4$. }}\label{table1}
\centering
\renewcommand\arraystretch{1.4}
\setlength{\tabcolsep}{6mm}{
\begin{tabular}{|c|c|}
\hline
Index Bits & Per-chirp Sequence ${\bf c}_{2,g}$ \\ \hline \hline
{[0~0]}  & ${\bf c}_{2,g}^{1}=[{c}_{2,g,0}^1,{c}_{2,g,1}^1,{c}_{2,g,2}^1,{c}_{2,g,3}^1]$           \\ \hline
{[0~1]}  & ${\bf c}_{2,g}^{2}=[{c}_{2,g,0}^2,{c}_{2,g,1}^2,{c}_{2,g,2}^2,{c}_{2,g,3}^2]$           \\ \hline
{[1~0]}  & ${\bf c}_{2,g}^{3}=[{c}_{2,g,0}^3,{c}_{2,g,1}^3,{c}_{2,g,2}^3,{c}_{2,g,3}^3]$           \\ \hline
{[1~1]}  & ${\bf c}_{2,g}^{4}=[{c}_{2,g,0}^4,{c}_{2,g,1}^4,{c}_{2,g,2}^4,{c}_{2,g,3}^4]$           \\ \hline
\end{tabular}}
\vspace{-2mm}
\end{table}

\subsection{Channel Model}
To characterize the doubly selective channel behavior in high-mobility scenarios, we consider a generalized $P$-path LTV channel, where each generalized path is jointly characterized by a delay-Doppler pair rather than by the propagation delay tap alone. Specifically, the $p$-th generalized path is represented by the triplet  $\left(h_{{\rm \xi},p},l_{{\rm \xi},p},{\alpha}_{{\rm \xi},p}\right)$, where $h_{{\rm \xi},p}$, $l_{{\rm \xi},p}\in[0,l_{\rm max}]$, and ${\alpha}_{{\rm \xi},p}\in[-{\alpha}_{\rm max}, {\alpha}_{\rm max}]$ denote its complex gain, normalized delay, and normalized Doppler shift, respectively.\footnote{The physical Doppler frequency of the $p$-th path is given by
$\nu_{\xi,p}=v_{\xi}f_c\cos\theta_{\xi,p}/c_0$, where $v_{\xi}$ is the
relative mobile speed, $f_c$ is the carrier frequency, $c_0$ is the
speed of light, and $\theta_{\xi,p}$ is the angle between the direction
of motion and the $p$-th propagation path. Accordingly, the normalized
Doppler shift is
$\alpha_{\xi,p}=\nu_{\xi,p}/\Delta f
=v_{\xi}f_c\cos\theta_{\xi,p}/(c_0\Delta f)$.
The maximum Doppler shift corresponds to
$|\cos\theta_{\xi,p}|=1$, yielding
$\nu_{\max}=v_{\max}f_c/c_0$ and
$\alpha_{\max}=\nu_{\max}/\Delta f
=v_{\max}f_c/(c_0\Delta f)$.}
In practice, the normalized Doppler is usually fractional due to poor Doppler resolution caused by the limited AFDM frame length. Hence, ${\alpha}_{{\rm \xi},p}={a}_{{\rm \xi},p}+{\beta}_{{\rm \xi},p}$, where ${a}_{{\rm \xi},p}\in{\mathbb{Z}}$ and ${\beta}_{{\rm \xi},p}\in[-1/2,1/2)$ denote the integer and fractional Doppler components, respectively.
After removing the CPP, the channel matrix ${\bf H}_{\xi}$ in the time domain is given by
\begin{align}\label{AFDM.3}
{\bf H}_{\xi}
=
\sum_{p=0}^{P-1}
h_{\xi,p}
{\bf\Gamma}_{{\rm CPP}_{\xi,p}}
{\bf\Delta}_{\alpha_{\xi,p}}
{\bf\Pi}^{l_{\xi,p}}.
\end{align}
Here, $\mathbf{\Gamma}_{\mathrm{CPP},\xi,p}=\operatorname{diag}
\left(
\gamma_{\xi,p}[0],\ldots,\gamma_{\xi,p}[N-1]
\right)$ is the CPP-induced
phase-correction matrix, 
where
\begin{align}\label{AFDM.8}
\gamma_{\xi,p}[n]
\!=\!
\begin{cases}
e^{
-j2\pi c_1
\left[N^2-2N\left(l_{\xi,p}-n\right)\right]},~n<l_{\xi,p},\\[1mm]
1,~n\geq l_{\xi,p}.
\end{cases}
\end{align}
Moreover, $\mathbf{\Delta}_{\alpha_{\xi,p}}$ denotes the Doppler-shift
matrix, given by
\begin{align}\label{AFDM.9}
\mathbf{\Delta}_{\alpha_{\xi,p}}
=
\operatorname{diag}
\left(
1,
e^{-j2\pi\alpha_{\xi,p}/N},
\ldots,
e^{-j2\pi\alpha_{\xi,p}(N-1)/N}
\right),
\end{align}
and $\mathbf{\Pi}^{l_{\xi,p}}$ represents the cyclic delay-shift
operator associated with the normalized delay $l_{\xi,p}$, where
\begin{align}\label{AFDM.10}
\mathbf{\Pi}
=
\begin{bmatrix}
0      & \cdots & 0      & 1\\
1      & \cdots & 0      & 0\\
\vdots & \ddots & \ddots & \vdots\\
0      & \cdots & 1      & 0
\end{bmatrix}_{N\times N}.
\end{align}
Accordingly, after applying the DAFT at receiver $\xi$, the equivalent
DAFT-domain channel matrix is constructed as
$\mathbf{H}_{\mathrm{eff}}^{\xi}
=
\mathbf{\Lambda}_{c_2^\xi}
\mathbf{F}_{N}
\mathbf{\Lambda}_{c_1}
\mathbf{H}_{\xi}
\mathbf{\Lambda}_{c_1}^{H}
\mathbf{F}_{N}^{H}
\mathbf{\Lambda}_{c_2^\xi}^{H}.$
With a properly selected post-chirp parameter $c_1$,
$\mathbf{H}_{\mathrm{eff}}^{\xi}$ exhibits a structured sparse
representation in the DAFT domain, thereby facilitating path separation
and low-complexity detection in high-mobility scenarios.
\subsection{Receiver of the Proposed AFDM-PCS-IM}
At the receiver, after removing the CPP, the received signal at Bob or Eve in the time domain can be written as
\begin{align}\label{R-bob}
{\bf r}_{\rm \xi}={\bf H}_{\rm \xi}{\bf s}+{\bf n}_{\rm \xi},
\end{align}
where $\xi\in\{\rm B, E\}$, and ${\bf n}_{\xi}\sim\mathcal{CN}({\bf 0}_N, N_0{\bf I}_N)$ is the complex additive white Gaussian noise (AWGN) vector. Especially, $\xi={\rm B}$ represents the Alice $\rightarrow$ Bob link, and $\xi={\rm E}$ represents the Alice $\rightarrow$ Eve link. Then, the DAFT domain received signal can be obtained by perform the DAFT on ${\bf r}_{\rm \xi}$ as follows
\begin{align}\label{R-bob-daft}
{\bf y}_{\rm \xi}&={\bm\Lambda}_{c_2^{{\rm \xi}}}{ {\bf F}_{N}}{\bm \Lambda}_{{c_{1}}}{\bf H}_{\rm \xi}{{\bm\Lambda}_{c_1}^H{ {\bf F}_{N}^H}{\bm \Lambda}_{{c_{2}^{{\rm A}}}}^H{\bf x}}+\underbrace{{\bm\Lambda}_{c_2^{{\rm \xi}}}{ {\bf F}_{N}}{\bm \Lambda}_{{c_{1}}}{\bf n}_{\rm \xi}}_{{{\bf {\overline n}}_{\rm \xi}}}\nonumber\\
&=\underbrace{{\bm\Lambda}_{c_2^{{\rm \xi}}}{ {\bf F}_{N}}{\bm \Lambda}_{{c_{1}}}{\bf H}_{\rm \xi}{{\bm\Lambda}_{c_1}^H{ {\bf F}_{N}^H}{\bm\Lambda}_{c_2^{{\rm \xi}}}^{H}}}_{{\bf H}_{\rm eff}^{\rm \xi}}{\bm\Lambda}_{c_2^{{\rm \xi}}}{\bm \Lambda}_{{c_{2}^{{\rm A}}}}^H{\bf x}+{{\bf {\overline n}}_{\rm \xi}}\nonumber\\
&={\bf H}_{\rm eff}^{\rm \xi}{\bm \Lambda}_{{c_{2}^{{\rm \xi}}}}{\bm \Lambda}_{{c_{2}^{{\rm A}}}}^H{\bf x}+{{\bf {\overline n}}_{\rm \xi}},
\end{align}
where ${\bf c}_{2}^{{\rm \xi}}$ represents the pre-chirp parameter sequence used by Bob or~Eve.
{To recover the transmitted bits, we first perform channel equalization using the minimum mean square error (MMSE) detector, which can be given by
\begin{align}\label{MMSE}
{\widehat{\bf x}^{\rm \xi}}_{\rm {MMSE}}=
{{\bf H}_{\rm eff}^{{\rm \xi}H}}
\left(
{{\bf H}_{\rm eff}^{{\rm \xi}}}
{{\bf H}_{\rm eff}^{{\rm \xi}H}}+
\frac{1}{\gamma_s}{{\bf I}_{N}}
\right)^{-1}
{\bf y}_{\rm \xi},
\end{align}
where ${\gamma_s}={E}_{s}/{N}_{0}$, and ${E}_{s}$ denotes the average power per symbol.
Then, the maximum-likelihood (ML) detection processes at Bob and Eve of the $g$-th group, respectively, can be formulated~as
\begin{align}\label{ML-bob}
  \left\{{\hat{\bf x}}_{g},{\hat {\bf c}_{2,g}^{{\rm A}}} \right\}_{\rm Bob}
  = \arg\!\!\!
  \underset{{ \substack{
      {\bf c}_{2,g}^{\rm A}\in {\mathcal C}_{2,g},
      {\bf x}_{g}\in{\mathcal{X}}
  }}}{\min}\!\!
  \left\| {\widehat{\bf x}}_{{\rm MMSE},g}^{\rm B}\!
    -\!
     {\bm \Lambda}_{{c_{2,g}^{{\rm B}}}}
     {\bm \Lambda}_{{c_{2,g}^{{\rm A}}}}^{H}
     {\bf x}_{g}
  \right\|^2,
\end{align}
and
\begin{align}\label{ML-eve}
\left\{{\hat{\bf x}}_{g},{\hat {\bf c}_{2,g}^{{\rm A}}} \right\}_{\rm Eve}
  = \arg\!\!\!
  \underset{{ \substack{
      {\bf c}_{2,g}^{\rm A}\in {\mathcal C}'_{2,g},
      {\bf x}_{g}\in{\mathcal{X}}
  }}}{\min}\!\!
  \left\| {\widehat{\bf x}}_{{\rm MMSE},g}^{\rm E}\!
    -\!
     {\bm \Lambda}_{{c_{2,g}^{{\rm E}}}}
     {\bm \Lambda}_{{c_{2,g}^{{\rm A}}}}^{H}
     {\bf x}_{g}
  \right\|^2,
\end{align}
where ${\bf c}_{2,g}^{\rm B}\in\mathcal{C}_{2,g}$ and ${\bf c}_{2,g}^{\rm E}\in\mathcal{C}'_{2,g}$ denote the pre-chirp parameter sequences employed at the receivers of Bob and Eve, respectively. ${\mathcal{C}_{2,g}}$, ${\mathcal{C}^{'}_{2,g}}$ and $\mathcal{X}$ represent the codebook set of the $g$-th group pre-chirp parameters for Bob, Eve and the set of constellations for the modulated bits, respectively.
}

{\bf {Remark 2:}}
According to~\eqref{ML-bob} and~\eqref{ML-eve}, one can observe that the pre-chirp parameter codebook ${\mathcal{C}_{2,g}}$ plays a dual role in the proposed AFDM-PCS-IM system. On one hand, over the legitimate Alice $\rightarrow$ Bob link, Bob performs ML detection by jointly searching the transmitted symbols ${\bf x}_{g}$ and the pre-chirp parameter sequence ${\bf c}_{2,g}^{\rm A}$ within the codebook ${\mathcal{C}_{2,g}}$.
Therefore, how distinguishable the different pre-chirp parameter sequences are under the legitimate channel determines the detection reliability at Bob.
On the other hand, although Eve can also perform channel estimation and reconstruct the DAFT-domain channel matrix, the effective pre-chirp parameter codebook ${\mathcal{C}'_{2,g}}$ observed over the Alice $\rightarrow$ Eve link is uncorrelated with the legitimate pre-chirp parameter codebook ${\mathcal{C}_{2,g}}$, i.e., ${\mathcal{C}_{2,g}} \cap {\mathcal{C}'_{2,g}} \approx \bm\emptyset$. As a result, Eve fails to correctly recover the transmitted bits.

\section{pre-chirp parameter codebook Design and Encryption Principles}\label{codebook}
According to the discussion in {\bf Remark 2}, we observe that a high-performance codebook ${\mathcal C}_{2,g}$ should satisfy both reliability and secrecy requirements. Therefore, we divide the codebook design into two parts: (i) constructing a reliability-oriented codebook that maximizes the distinguishability of pre-chirp sequences at Bob, and (ii) incorporating CSI-driven perturbations to enhance secrecy without destroying the orthogonality structure required by Bob.

\subsection{Reliability Requirement and Orthogonal Codebook Construction}
By analyzing the ML detection metric in \eqref{ML-bob}, the pairwise error probability (PEP) is minimized when the effective matrices ${\bm \Lambda}_{{c}_{2,g}^{\rm B}}{\bm \Lambda}_{{c}_{2,g}^{i}}^{H}$ and ${\bm \Lambda}_{{c}_{2,g}^{\rm B}}{\bm \Lambda}_{{c}_{2,g}^{j}}^{H}$ are mutually uncorrelated or ideally orthogonal, where $\forall {{\bf c}_{2,g}^{i}},~{{\bf c}_{2,g}^{j}}\in{\mathcal{C}_{2,g}}$.
This motivates {\bf Design Criterion~1}, which requires that
\begin{align}\label{Design-Criterion}
\Big\langle
{\bm \Lambda}_{{c}_{2,g}^{i}},
{\bm \Lambda}_{{c}_{2,g}^{j}}
\Big\rangle \approx 0, \qquad i\neq j,
\end{align}
where $\langle \mathbf{A}, \mathbf{B} \rangle = \mathrm{tr}(\mathbf{A}^{H} \mathbf{B})$ denotes the Frobenius inner product between two matrices. To satisfy this criterion, we construct ${\mathcal C}_{2,g}$ based on orthogonal sequence sets.
Without loss of generality, the Walsh-Hadamard code is particularly appealing due to its perfect orthogonality and constant amplitude structure, i.e.,\footnote{Actually, owing to the flexible configurability of the pre-chirp parameters, any prescribed constant-modulus phase codebook can be realized in the pre-chirp parameter domain. Specifically, for a target entry $u_k[m]=e^{j\phi_k[m]}$, the corresponding pre-chirp parameter can be selected as $c_{2,g,\mathrm{orth}}^{k}[m]=-\phi_k[m]/(2\pi f(g,m)^2)~(\bmod~1)$ for $f(g,m)\neq0$.
Therefore, other orthogonal codebooks, such as DFT and complex Hadamard codebooks, can also be employed. Low-correlation codebooks are
also applicable, although their residual correlation may reduce the distinguishability between different index patterns. The
Walsh-Hadamard codebook is adopted here because its binary entries
enable a simple $\{0,\pi\}$ phase mapping and its structured
representation facilitates the fast Walsh-Hadamard transform and the
proposed Walsh-constrained detector.}
\begin{align}\label{Walsh-code}
\Big\langle
{\bf w}_{i},
{\bf w}_{j}
\Big\rangle = 0, \qquad i\neq j.
\end{align}
where ${\bf w}_{i}$ and ${\bf w}_{j}$ represent the Walsh-Hadamard sequences.
Then, for the $k$-th  Walsh-Hadamard sequence $\mathbf{w}_k$, the corresponding pre-chirp
parameter sequence $\mathbf c_{2,g,\rm orth}^{k}$ can be obtained through the following
mapping.

Let $\mathbf{w}_k = [\,w_k[0], w_k[1], \ldots, w_k[M-1]\,]^{T}$
denote the $k$-th orthogonal sequence in the Walsh-Hadamard matrix
$\mathbf{W}_{M}$.
The relationship between $\mathbf{w}_k$ and the pre-chirp parameter matrix ${\bm \Lambda}_{{c}_{2,g,\rm orth}^{k}}$ can be expressed as
\begin{align}\label{walsh-matching}
{\Lambda}_{c_{2,g,\rm orth}^{k}}[m,m]
=e^{-j2\pi c_{2,g,\rm orth}^{k}[m]f(g,m)^2}
=
w_k[m].
\end{align}
where $f(g,m)=(g-1)M+m$.
Since each Walsh-Hadamard symbol satisfies $w_k[m]\in\{+1,-1\}$, the pre-chirp parameter can be explicitly
derived as
\begin{equation}
c_{2,g,\rm orth}^{k}[m]
=  \frac{\arg(w_k[m])}{2\pi f(g,m)^2}~~~(\bmod~1).
\label{eq:c2-mapping}
\end{equation}
For the zeroth chirp subcarrier, we define
$c_{2,1,\rm orth}^{k}[0] \triangleq 0$ to avoid the singularity caused by
the factor $1/n^2$.
This mapping converts the binary orthogonal Walsh sequences into
chirp-domain phase parameters while preserving their mutual orthogonality.

\subsection{Secrecy Requirement and CSI-Driven Perturbation Design}
Although the orthogonal pre-chirp parameter sequences ensure reliable detection at Bob, this structure alone does not guarantee secrecy. To strengthen the resilience against eavesdropping of the proposed AFDM-PCS-IM scheme, we introduce an additional CSI-driven perturbation that create uncertainty for Eve during the detection process.
{Specifically, for each pre-chirp parameter, we add a perturbation term of the form
\begin{align}\label{eq:csi-perturbation}
c_{2,\mathrm{per}}[n]
   = \frac{s[n]}{2n^2}\, Q\!\left( \left|\mu[n]\right|\right),
\end{align}
where $s[n]\in\{\pm1\}$ is a sign sequence derived from the legitimate CSI. 
To this end, we first define an auxiliary real-valued CSI metric for the $n$-th chirp subcarrier as
\begin{align}\label{eq:csi-metric}
\mu[n] \triangleq 
\Re\left\{\sum_{j=1}^{N} H_{\xi}^{\mathrm{Fre}}[j,n] \right\},
\end{align}
where $\mathbf{H}_{\xi}^{\mathrm{Fre}}=\mathbf{F}_{N}\mathbf{H}_{\xi}\mathbf{F}_{N}^{H}$ collects the 
frequency-domain channel taps associated with all chirp subcarriers and 
$\Re\{\cdot\}$ denotes the real-part operator. The sign sequence is then given by
$
s[n] = \mathrm{sign}\big(\mu[n]\big)$, 
and $Q(\cdot)$ denotes a scalar quantization function to improve robustness of the proposed system caused by imperfect CSI.} 
The detail of quantization process will be discussed in Section~\ref{quantization}.
Crucially, a fixed public baseline $c_{2,0}$ and the CSI-driven perturbation $c_{2,\rm per}[n]$ are commonly applied to all codewords, i.e.,
\begin{align}\label{lemma-1}
c_{2}^{k}[n]
=
c_{2,0}
+
c_{2,\rm orth}^{k}[n]
+
c_{2,\rm per}[n],
\quad \forall k.
\end{align}
Here, $c_{2,0}$ is a fixed and public irrational baseline selected
offline, after the finite Walsh codebook and quantizer output alphabet
have been specified, such that every resulting total pre-chirp
parameter is irrational. Accordingly, the resulting pre-chirp
parameters satisfy the irrational-number branch of the conventional
AFDM design criterion. Since $c_{2,0}$ and $c_{2,\rm per}[n]$ are common to all codewords, it
only contributes a common unitary phase factor and therefore preserves
the pairwise codebook orthogonality established in
Lemma~\ref{lem:orth-preserve}.
The perturbation depends on the legitimate CSI, so that Eve cannot
reproduce the same structure.
%
For clarity, the complete procedure from Walsh-Hadamard sequence
selection to CSI-driven perturbation and encrypted pre-chirp codebook
construction is summarized in Algorithm~\ref{alg:encrypted-codebook}.
\begin{algorithm}[t]
\caption{CSI-Driven Encrypted Pre-Chirp Codebook Construction}
\label{alg:encrypted-codebook}
\begin{algorithmic}[1]
\Require Group size $M$, number of groups $G$, number of codewords
$K$, legitimate CSI $\mathbf{H}_{\mathrm{B}}^{\mathrm{Fre}}$, and
quantizer $Q(\cdot)$.
\Ensure Encrypted pre-chirp codebooks
$\{\mathcal{C}_{2,g}\}_{g=1}^{G}$.

\State Generate the Walsh--Hadamard matrix $\mathbf{W}_M$ and select
$K$ mutually orthogonal sequences
$\{\mathbf{w}_k\}_{k=1}^{K}$.

\For{$g=1,\ldots,G$}
    \State Map the selected Walsh sequences to the orthogonal
    pre-chirp parameter sequences
    $\{\mathbf{c}_{2,g,\mathrm{orth}}^{k}\}_{k=1}^{K}$
    according to \eqref{eq:c2-mapping}.

    \State Calculate the legitimate CSI metric $\mu[n]$ and sign
    sequence $s[n]$ according to \eqref{eq:csi-metric}.

    \State Generate the CSI-driven perturbation sequence
    $\mathbf{c}_{2,\mathrm{per}}$ according to
    \eqref{eq:csi-perturbation}, with
    $c_{2,\mathrm{per}}[0]\triangleq0$.

    \State Add the public irrational baseline and the common CSI-driven
perturbation to each orthogonal pre-chirp sequence according to
\eqref{lemma-1}.

    \State Construct
    $\mathcal{C}_{2,g}
    =
    \{\mathbf{c}_{2,g}^{1},\ldots,\mathbf{c}_{2,g}^{K}\}$.
\EndFor

\State \Return $\{\mathcal{C}_{2,g}\}_{g=1}^{G}$.
\end{algorithmic}
\end{algorithm}

\begin{lemma}\label{lem:orth-preserve}
The common baseline $c_{2,0}$ and the CSI-driven perturbation
$c_{2,\rm per}[n]$ introduced in \eqref{lemma-1} do not destroy the
mutual orthogonality of the reliability-oriented codebook
$\mathcal{C}_{2,g}$ constructed from Walsh-Hadamard sequences.
\end{lemma}
\begin{proof}
Let $\mathbf{\Lambda}_{c_{2,g,\rm orth}^{k}}$ denote the diagonal
pre-chirp parameter matrix generated from the $k$-th Walsh-Hadamard
sequence. Define the common phase matrix of the $g$-th group as
\begin{align}
\mathbf{\Lambda}_{{\rm com},g}
\triangleq &
\mathrm{diag}\!\Big(
e^{-j2\pi \widetilde{c}_{2}[f_0^g](f_0^g)^2},
\ldots,
e^{-j2\pi \widetilde{c}_{2}[f_m^g](f_m^g)^2},
\nonumber\\
&
\ldots,
e^{-j2\pi \widetilde{c}_{2}[f_{M-1}^g](f_{M-1}^g)^2}
\Big),
\end{align}
where $\widetilde{c}_{2}[n]
\triangleq
c_{2,0}+c_{2,\rm per}[n]$, and $f_m^g\triangleq f(g,m)=(g-1)M+m$.
Since $c_{2,0}\in\mathbb{R}$ and
$c_{2,\rm per}[n]\in\mathbb{R}$,
$\mathbf{\Lambda}_{{\rm com},g}$ is unitary, i.e.,
\begin{align}
\mathbf{\Lambda}_{{\rm com},g}^{H}
\mathbf{\Lambda}_{{\rm com},g}
=
\mathbf{I}_M.
\end{align}
With the same baseline and perturbation applied to all pre-chirp
sequences, the resulting pre-chirp matrix satisfies
\begin{align}
\mathbf{\Lambda}_{c_{2,g}^{k}}
=
\mathbf{\Lambda}_{{\rm com},g}
\mathbf{\Lambda}_{c_{2,g,\rm orth}^{k}}.
\end{align}
Therefore, for any pre-chirp sequence, i.e., $\forall \mathbf{c}_{2,g}^{i}, \mathbf{c}_{2,g}^{j} \in \mathcal{C}_{2,g}$ with $i\neq j$, we have
\begin{align}
\mathbf{\Lambda}_{c_{2,g}^{i}}
\mathbf{\Lambda}_{c_{2,g}^{j}}^{H}
=
\mathbf{\Lambda}_{{\rm com},g}
\mathbf{\Lambda}_{c_{2,g,\rm orth}^{i}}
\mathbf{\Lambda}_{c_{2,g,\rm orth}^{j}}^{H}
\mathbf{\Lambda}_{{\rm com},g}^{H}.
\end{align}
Since the trace is invariant under unitary similarity transformations, we can obtain
\begin{align}
\Big\langle
\mathbf{\Lambda}_{c_{2,g}^{i}},
\mathbf{\Lambda}_{c_{2,g}^{j}}
\Big\rangle
=
\Big\langle
\mathbf{\Lambda}_{c_{2,g,\rm orth}^{i}},
\mathbf{\Lambda}_{c_{2,g,\rm orth}^{j}}
\Big\rangle.
\end{align}
We can find that the pairwise inner products of any pre-chirp matrix remain unchanged after perturbation.
Because the original Walsh-based codebook satisfies
\begin{align}
\big\langle
\mathbf{\Lambda}_{c_{2,g,\rm orth}^{i}},
\mathbf{\Lambda}_{c_{2,g,\rm orth}^{j}}
\big\rangle=0,\quad i\neq j,
\end{align}
the perturbed pre-chirp parameter codebook preserves orthogonality.
\end{proof}

\section{Robustness Mechanism Based on CSI Quantization}\label{quantization} Owing to channel noise and hardware limitations, it is difficult to obtain perfect CSI in practice scenario. If the perturbation term is directly generated by the unquantized CSI, the inconsistency between Alice and Bob's estimated CSI may cause mismatched chirp parameter codebooks, thereby degrading the transmission reliability.
In this section, we consider the design of a quantizer under imperfect CSI. 
Without loss of generality, we consider the imperfect CSI amplitude
\begin{equation}
{\hat a}[n] = a[n] + \epsilon[n],
\end{equation}
where $a[n]=|\mu[n]|$ denotes the magnitude of the true CSI metric,
$\hat a[n]=|\hat{\mu}[n]|$ represents its estimated magnitude, and
$\epsilon[n]\sim\mathcal{N}(0,\sigma_h^2)$ denotes the corresponding
estimation error.

To address the codebooks mismatch caused by the CSI disparity between Alice and Bob, we develop a noise-aware CSI quantizer
$Q(\cdot)$ applied to $\hat a[n]$.
The quantizer utilizes two main parameters: a dead-zone width $d$ and a step size $\Delta$, both are related to the estimation error variance  $\sigma_h$. The formulas are given as
$d = k\,\sigma_h$ and
$\Delta = \gamma\,\sigma_h$
where $k>0$ and $\gamma>0$ are two adjustable parameters that determine the resolution of the quantizer.
Then, we divide the CSI amplitude range outside the dead-zone into $L$ non-zero quantization intervals. The output of the quantizer for the $i$-th interval $q_i$ and the corresponding decision thresholds $ b_i$ are then given by
\begin{align}\label{q-19}
q_i = d + \Big(i-\tfrac{1}{2}\Big)\alpha, 
\end{align}
and
\begin{align}\label{q-20}
b_i = d + i\Delta,
\end{align}
where $i=0,\dots,L$, and $q_0=0$. Here, $\Delta$ specifies the width of the input decision intervals,
which governs the tolerance to CSI estimation mismatch. $\alpha$
sets the separation between nonzero output labels and thereby controls
the perturbation strength.

Therefore, the quantizer output $Q(\hat a[n])$ is then defined as
\begin{equation}
q[n]=
\begin{cases}
0, & 0 \le \hat a[n] < b_0,\\[2pt]
q_i, & b_{i-1} \le \hat a[n] < b_i,\\
q_L, & \hat a[n] \ge b_{L-1}.
\end{cases}
\label{eq:CSI-quantizer}
\end{equation}
The zero output within the dead zone is intentionally retained to
suppress unreliable low-magnitude CSI components, whose signs are
particularly sensitive to channel estimation errors. Thus, an
individual $Q(|\hat{\mu}[n]|)=0$ is allowed and simply gives
$c_{2,\mathrm{per}}[n]=0$ on the corresponding chirp subcarrier. To
exclude the degenerate case in which all CSI components are quantized
to zero, the current CSI estimate is discarded and channel estimation
is repeated if $\max_{1\leq n\leq N-1}|\hat{\mu}[n]|<d$.
This degenerate event can occur only when all $N-1$ CSI metrics
simultaneously fall within the dead zone, which is highly unlikely
under practical operating conditions. Therefore, repeating the channel estimation procedure only in
such rare cases is not expected to introduce appreciable average
latency or protocol overhead, nor to become a bottleneck for real-time
transmission.
With this construction, Alice and Bob generate identical quantized CSI
values when their local CSI estimates fall within the same decision
interval and have the same sign,~i.e.,
\begin{equation}
b_{i-1}\leq
\hat a_{\rm A}[n],\hat a_{\rm B}[n]
<b_i,
\quad
Q(\hat a_{\rm A}[n])
=
Q(\hat a_{\rm B}[n]).
\end{equation}
Therefore, the perturbation term $c_{2,\rm per}[n]$ constructed from
the quantized CSI metric $Q(|\hat{\mu}[n]|)$ is identical at Alice and
Bob, which ensures that both parties generate the same encrypted
pre-chirp parameter codebook $\mathcal{C}_2$ under imperfect CSI
conditions.
Finally, the quantization process is summarized in
Algorithm~\ref{quantization-algorithm}.

%
\begin{algorithm}[t]
\caption{Noise-Aware CSI Quantization}\label{quantization-algorithm}
\begin{algorithmic}[1]
\Require Estimated CSI-metric magnitude
$\hat a[n]=|\hat{\mu}[n]|$;
        noise variance $\sigma_h^2$; quantizer parameters $k>0$, $\gamma>0$, and number of non-zero levels $L$.
\Ensure Quantized output $q[n]$.

\State $d \gets k\sigma_h,\ \ \Delta \gets \gamma\sigma_h$
\State $b_0 \gets d$
\For{$i = 1$ to $L$}
  \State $b_i \gets d + i\Delta$
  \State $q_i \gets d + (i-\tfrac{1}{2})\alpha$
\EndFor

\State \textbf{Quantization:}
\If {$0 \le \hat a[n] < b_0$}
  \State $q[n] \gets 0$
\Else
  \If {$\hat a[n] < b_{L-1}$}
    \State Find $i \in \{1,\dots,L-1\}$ such that $b_{i-1} \le \hat a[n] < b_i$
    \State $q[n] \gets q_i$
  \Else
    \State $q[n] \gets q_L$
  \EndIf
\EndIf

\State \Return $q[n]$
\end{algorithmic}
\end{algorithm}

\section{Proposed Two-layer Message Passing Detector with Walsh Code Constraints}
In this section, we first derive an equivalent received signal model to facilitate the explanation of the proposed low-complexity TLMP-WC detector. Based on this model, we introduce the principle of the proposed TLMP-WC algorithm for large-dimensional data detection in the AFDM-PCS-IM scheme. Subsequently, we analyze the computational complexity of the proposed algorithm to demonstrate its effectiveness and feasibility.

\subsection{Equivalent Received Signal Model for Walsh-Constrained Detection}
As discussed in~\eqref{R-bob-daft}, the DAFT-domain received signal at Bob can be expressed as
\begin{equation}\label{R-bob-daft-eq}
\mathbf{y}_{\rm B}
=
\mathbf{H}_{\rm eff}^{\rm B}
\boldsymbol{\Lambda}_{c_2^{\rm B}}
\boldsymbol{\Lambda}_{c_2^{\rm A}}^{H}
\mathbf{x}
+ \overline{\mathbf{n}}_{\rm B},
\end{equation}
where $\mathbf{H}_{\rm eff}^{\rm B}$ denotes the DAFT-domain sparse channel matrix and
$\overline{\mathbf{n}}_{\rm B}\sim\mathcal{CN}(\mathbf{0}_{N},N_0\mathbf{I}_{N})$ represents the equivalent noise vector at~Bob.
According to the Walsh-Hadamard based design of the pre-chirp parameter set $\mathcal{C}_2$, the product
$\boldsymbol{\Lambda}_{c_2^{\rm B}} \boldsymbol{\Lambda}_{c_2^{\rm A}}^{H}$
exhibits an inherent Walsh-constrained structure, i.e.,
\begin{align}
\boldsymbol{\Lambda}_{c_2^{\rm B}} \boldsymbol{\Lambda}_{c_2^{\rm A}}^{H}
=
\mathrm{diag}\{ {\bf w}_{k_1}^T,\ldots,{\bf w}_{k_g}^T,\ldots,{\bf w}_{k_G}^T \},
\end{align}
where ${\bf w}_{k_g} \in \mathbb{R}^{M\times 1}$ denotes the $k_g$-th column Walsh sequence selected from the Walsh-Hadamard matrix ${\bf{W}}_M$, and $k_g\in\{1,\ldots,M\}$ denotes the $g$-th group activated index.
By explicitly exploiting this structure, \eqref{R-bob-daft-eq} can be reformulated as an equivalent model
\begin{align}\label{eq-24}
\mathbf{y}_{\rm B}&
=
\mathbf{H}_{\rm eff}^{\rm B}
\mathbf{W}_{\rm eff}
\mathbf{x}'
+ \overline{\mathbf{n}}_{\rm B}\nonumber\\
&=\mathbf{H}_{\rm eff}^{\rm B}
\mathbf{z}
+ \overline{\mathbf{n}}_{\rm B},
\end{align}
where
$\mathbf{W}_{\rm eff} = \mathrm{diag}\{\mathbf{W}_{M}, \mathbf{W}_{M}, \ldots, \mathbf{W}_{M}\}\in\mathbb{R}^{N \times N}$
is a block-diagonal matrix composed of $G$ Walsh-Hadamard matrices
$\mathbf{W}_{M} \in \mathbb{R}^{M \times M}$, and $\mathbf{z}=\mathbf{W}_{\rm eff}
\mathbf{x}'$.
In this equivalent representation, the transmitted signal $\mathbf{x}'=[{\mathbf{x}'_1}^T,\ldots,{\mathbf{x}'}_G^T]^T\in\mathbb{C}^{N \times 1}$ is characterized by a \emph{block-sparse structure}. 
Specifically, the $g$-th sub-vector denotes as $\mathbf{x}'_{g}= x_g \mathbf{e}_{k_g}\in\mathbb{C}^{M \times 1}$, and $\mathbf{e}_{k_g}\in\mathbb{R}^{M\times1}$ denotes the $k_g$-th canonical basis vector, i.e., $k_g$-th element equals one and all the others are zero.
It is worth noting that, since $\mathbf{z}$ is obtained from $\mathbf{x}'$ through the Walsh-Hadamard transformation $\mathbf{W}_{\rm eff}$, the elements of $\mathbf{z}$ also take values from the same symbol alphabet $\mathcal{S}$.
Although the optimal maximum a posteriori (MAP) detector achieves the best performance under the sparse Bayesian model, its computational complexity increases exponentially with the signal dimension $N$, rendering it unsuitable for large-scale detection and applications. According to~\eqref{eq-24}, we can give a low-complexity suboptimal MAP detector. Since $\mathbf{z}=\mathbf{W}_{\rm eff}
\mathbf{x}'$ is a deterministic one-to-one transformation, the joint posterior ${\rm Pr}(\mathbf{x}',\mathbf{z}\mid \mathbf{y}_{\rm B})$ is equivalent to ${{\rm Pr}({\bf z}\mid{\bf y}_{\rm B})}$. In particular, the MAP of ${{\rm Pr}({\bf z}\mid{\bf y}_{\rm B})}$ can be approximated~as 
\begin{align}
{\rm Pr}(\mathbf{x}',\mathbf{z}\mid \mathbf{y}_{\rm B})\!
\propto\!
{\rm Pr}(\mathbf{y}_{\rm B}\mid \mathbf{z})
\prod_{c=1}^{N} {\rm Pr}\big(z(c)\mid \mathbf{x}'\big)
\prod_{g=1}^{G} {\rm Pr}(\mathbf{x}'_g),
\end{align}
where the likelihood factor ${\rm Pr}(\mathbf{y}_{\rm B}\mid \mathbf{z})$ corresponds to the
linear channel observation model, the mapping factors ${\rm Pr}\big(z(c)\mid \mathbf{x}'\big)$
represent the deterministic Walsh-Hadamard spreading relationship between
$\mathbf{x}'$ and $\mathbf{z}$, and the prior factors ${\rm Pr}(\mathbf{x}'_g)$ characterize
the Walsh-constrained block-sparse structure by restricting each group to a single
valid activation pattern.
This Walsh-constrained block-sparse formulation significantly reduces the effective search space of the detection problem, thereby laying the foundation for the development of a low-complexity two-layer message passing detector.

\begin{figure}[t]
	\center
	\includegraphics[width=3.6in,height=1.4in]{{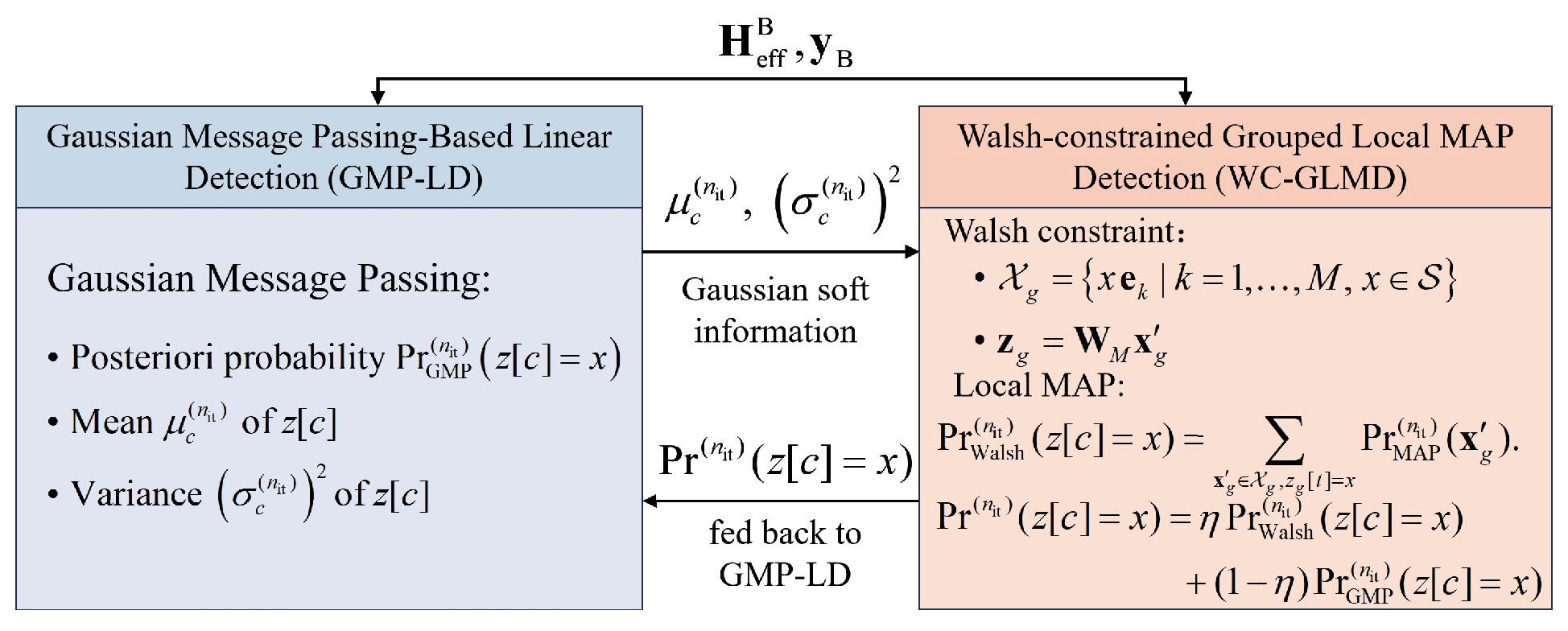}}
	\vspace{-0.0cm}
	\caption{{Block diagram of the iterative process in the proposed TLMP-WC detector.}}
	\label{graphical_factor}  
	\vspace{-4mm}
\end{figure}
\subsection{Proposed TLMP-WC Detector}
Fig.~\ref{graphical_factor} illustrates the block diagram of the iterative process in the proposed TLMP-WC detector.
For large-dimensional signal detection, performing accurate message passing across the entire graph would result in excessive computational overhead.
Therefore, we adopt a two-layer approximate Bayesian inference framework. 
Specifically, the first layer corresponds to a linear detector based on Gaussian message passing (GMP-LD), which consists of the observation nodes $y[d]$ and the variable nodes $z[c]$.
The second layer performs Walsh-constrained grouped local maximum likelihood detection (WC-GLMD).
Through iterative information exchange between the two layers, a two-layer message-passing detector with Walsh code constraints is developed.
The detailed operations of the $n_{\rm it}$-th iteration are introduced as follows.

{{\bf {\em { 1)  Gaussian Message Passing-Based Linear Detection (GMP-LD):}}}}
In this layer, the posterior probability of each variable node ${z}[c]$ is calculated based on the prior information ${\rm Pr}^{(n_{\rm it}-1)}{(z[c]=x)}$ provided by the seconde layer from the pervious iteration. The initialization probability is set to ${\rm Pr}^{(0)}{(z[c]=x)}=1/|\mathcal{S}|$ in the first iteration. Then, according to the posterior probability, the mean and variance of  ${z}[c]$ can be obtained. Specifically, we can give the received signal of the $d$-th observation node as follows
\begin{align}\label{eq-26}
{y}[d]
=
{H}_{\rm eff}^{\rm B}[d,c]
{z}[c]+ 
\underbrace{\sum\limits_{e\in {\mathcal{R}[d]}, e\ne c}
{H}_{\rm eff}^{\rm B}[d,e]
{z}[e]
+ \overline{{n}}_{\rm B}[d]}_{\wp_{d,e}},
\end{align}
where ${\mathcal{R}[d]}$ denotes the set of non-zero element indices of the $d$-th row of $\mathbf{H}_{\rm eff}^{\rm B}$. According to the central limit theorem (CLT), the sum of contributions from the other variable nodes to $y[d]$ can be approximated as a Gaussian random variable, i,e., $\wp_{d,e} \sim \mathcal{CN}\!(\mu_{d,e},\, \sigma_{d,e}^2)$~\cite{tao_JSAC}. The corresponding mean and variance of the $n_{\rm it}$-th iteration can be calculated as follows
\begin{align}\label{eq-27}
\mu_{d,e}^{(n_{\rm it})}
=\sum\limits_{e\in {\mathcal{R}[d]}, e\ne c}
{H}_{\rm eff}^{\rm B}[d,e]
\sum\limits_{x\in {\mathcal {S}}}
x{\rm Pr}^{(n_{\rm it}-1)}{(z[e]=x)},
\end{align}
and
\begin{align}\label{eq-28}
   {{\left( \sigma _{d,e}^{({{n}_{\rm{it}}})} \right)}^{2}}=&\sum\limits_{e\in \mathcal{R}[d],e\ne c}
   \!\!\!{{{\left| {{H}^{\rm B}_{\rm{eff}}}[r,e] \right|}^{2}}}
   \left( \sum\limits_{x\in {{\mathcal{S}}}}
   {{{\left| x \right|}^{2}}}{\rm{Pr}}^{\left( {{n}_{\rm{it}}}-1 \right)}{(z[e]=x)}\right. \nonumber\\
 & \left.
 -{{\left| \sum\limits_{x\in {\mathcal {S}}}
x{\rm Pr}^{(n_{\rm it}-1)}{(z[e]=x)} \right|}^{2}} \right)
+{{N}_{0}}.
\end{align}
Then, when $z[c]=x$ occurs, the conditional likelihood probability of observation node $y[d]$ can be given by
\begin{align}\label{eq-29}
{{\rm{Pr}}\!\left({y[d]|{{z}[c]\!=\!x}}\right)}
\!\propto\!
{\rm{exp}}
\!\left(\!\frac{-\left|{{y}[d]}\!-\!
\mu_{d,e}^{({{n}_{\rm it}})}\!-\!
{H^{\rm B}_{\rm{eff}}[d,c]}x\right|^2}{{{\left( \sigma _{d,e}^{({{n}_{\rm{it}}})} \right)}^{2}}}
\!\right).
\end{align}
By collecting all observations $y(d)$ whose indices belong to $\mathcal L[c]$, i.e., the set of observation nodes connected to the $c$-th variable node, we compute the unnormalized probability estimation as
\begin{align}\label{eq-30}
{\rm Pr}_{\mathrm{GMP}}^{(n_{\rm it})}\bigl(z[c]=x\bigr)
= \prod_{d \in {\mathcal{L}[c]}} p\bigl(y(d)\mid z(c)=x\bigr),
\end{align}
where $\mathcal L[c]$ denotes the set of observation indices $d$ such that $H_{\mathrm{eff}}^{\mathrm B}[d,c] \neq 0$.
After normalization, we can obtain
\begin{align}\label{eq-31}
{\rm {Pr}}_{\mathrm{GMP}}^{(n_{\rm it})}\bigl(z[c]=x\bigr)
= \frac{{\rm{Pr}}_{\mathrm{GMP}}^{(n_{\rm it})}\bigl(z[c]=x\bigr)}
{\sum_{x' \in \mathcal{S}} {\rm{Pr}}_{\mathrm{GMP}}^{(n_{\rm it})}\bigl(z(c)=x'\bigr)}.
\end{align}
Based on the results of equation~\eqref{eq-31}, we can compute the mean $\mu_{c}^{(n_{\rm it})}$ and variance ${{\left( \sigma _{c}^{({{n}_{\rm{it}}})} \right)}^{2}}$ of $z[c]$, which will be transmitted to the second layer as prior information, i.e.,
\begin{align}\label{eq-32}
\mu_{c}^{(n_{\rm it})}=
\sum_{x \in \mathcal{S}} x 
{\rm {Pr}}_{\mathrm{GMP}}^{(n_{\rm it})}\left(z[c]=x\right),
\end{align}
and 
\begin{align}\label{eq-33}
{{\left( \sigma _{c}^{({{n}_{\rm{it}}})} \right)}^{2}}=
\sum_{x \in \mathcal{S}} |x|^2 
{\rm {Pr}}_{\mathrm{GMP}}^{(n_{\rm it})}\bigl(z[c]=x\bigr)
- \left|\mu_{c}^{(n_{\rm it})}\right|^2 .
\end{align}

{{\bf {\em { 2) Walsh-constrained Grouped Local MAP Detection (WC-GLMD):}}}}
{
In this layer, the Walsh-constrained block-sparse structure of $\mathbf{x}'$ is
explicitly exploited through grouped local MAP detection.
Conditioned on the symbol mean $\mu_{c}^{(n_{\rm it})}$ and variance ${{\left( \sigma _{c}^{{{n}_{\rm{it}}}} \right)}^{2}}$ obtained from the first layer at the
$n_{\rm it}$-th iteration, refined soft information of $z[c]$ is generated and subsequently
fed back to the first layer to form an iterative loop.

For the $g$-th group, due to the Walsh-constrained block-sparse structure, only one element in
$\mathbf{x}'_g$ is nonzero and must belong to the constellation set
$\mathcal{S}$. 
Hence, we define $\mathbf{x}'_g\in\mathcal{X}_g$, and the set of valid activation patterns is
\begin{equation}\label{eq-37}
\mathcal{X}_g
=
\left\{
x\,\mathbf{e}_k
\;\middle|\;
k=1,\ldots,M,\; x\in\mathcal{S}
\right\},
\end{equation}
where $\mathbf{e}_k$ denotes the $k$-th canonical basis vector of length $M$,
i.e., a vector whose $k$-th entry is one and all other entries are zero, and the length of the set is $M|\mathcal{S}|$.
Under $\mathbf{x}'_g \in \mathcal{X}_g$, the group symbol vector $\mathbf{z}_g$ is uniquely determined by the Walsh constrained
\begin{equation}\label{eq-38}
\mathbf{z}_g
=
\mathbf{W}_M \mathbf{x}'_g.
\end{equation}
Hence, the $g$-th group contribution to $y[d]$ can be represented as a deterministic function of $\mathbf{x}'_g$, i.e.,
\begin{equation}\label{eq-39}
s_{d,g}( \mathbf{x}'_g)=
\sum_{t \in \mathcal{G}_g \cap \mathcal{R}[d]}
H_{\mathrm{eff}}^{\mathrm{B}}[d,t]\,
z_g[t-(g-1)M],
\end{equation}
where $\mathcal{G}_g\triangleq\{(g-1)M+1,(g-1)M+2,\ldots,gM\}$, and $z_g[t-(g-1)M]$ can be determined according to the constraint in~\eqref{eq-38}.
Substituting $s_{d,g}( \mathbf{x}'_g)$ into the $g$-th group and the $d$-th observation node $y[d]$, the received signal can be
decomposed as
\begin{align}\label{eq-34}
y[d]
=
s_{d,g}( \mathbf{x}'_g)+
\underbrace{\sum_{e \in \mathcal{R}[d]\setminus\mathcal{G}_g}\!\!\!\!\!\!\!
H_{\mathrm{eff}}^{\mathrm{B}}[d,e]\, z[e]
+ \bar{n}_{\mathrm{B}}[d] }_{\phi_{d,g}},
\end{align}
where $\phi_{d,g}$ collects the contributions of symbols outside the $g$-th group and the observation noise.
Similar, we can approximate ${\phi_{d,g}}$ as a complex Gaussian variable ${\phi_{d,g}}\sim \mathcal{CN}\left(\mu_{d,g},
(\sigma_{d,g})^2\right)$. According to~\eqref{eq-32} and~\eqref{eq-33}, the mean and variance of ${\phi_{d,g}}$ in the $n_{\rm it}$-th iteration can be given~by
\begin{align}\label{eq-35}
\mu_{d,g}^{(n_{\mathrm{it}})}
&=
\sum_{e \in \mathcal{R}[d]\setminus\mathcal{G}_g}
H_{\mathrm{eff}}^{\mathrm{B}}[d,e]\,
\mu_e^{(n_{\mathrm{it}})},
\end{align}
and
\begin{align}\label{eq-36}
\left(\sigma_{d,g}^{(n_{\mathrm{it}})}\right)^2
&=
\sum_{e \in \mathcal{R}[d]\setminus\mathcal{G}_g}
\big|H_{\mathrm{eff}}^{\mathrm{B}}[d,e]\big|^2
(\sigma_e^{(n_{\mathrm{it}})})^2
+ N_0.
\end{align}
%
Then, based on the conditional Gaussian model, the log-likelihood ratio (LLR) of the candidate pattern $\mathbf{x}'_g$ is
\begin{equation}\label{eq-40}
\xi_g^{(n_{\mathrm{it}})}(\mathbf{x}'_g)
=
\sum_{d \in \mathcal{L}_g}
\frac{
-\big|y[d]-{\mu}_{d,g}^{(n_{\mathrm{it}})}-s_{d,g}( \mathbf{x}'_g)\big|^2
}{
(\sigma_{d,g}^{(n_{\mathrm{it}})})^2
},
\end{equation}
where the set of observation indices related to the $g$-th group is defined as
$\mathcal{L}_g
\triangleq
\big\{ d \,\big|\, \exists\, t \in \mathcal{G}_g,\;
H_{\mathrm{eff}}^{\mathrm{B}}[d,t] \neq 0 \big\}$.
The posterior probability of each activation pattern is then obtained by
normalizing the corresponding likelihood values as
\begin{equation}\label{eq-41}
\operatorname{Pr}_{\mathrm{MAP}}^{(n_{\mathrm{it}})}(\mathbf{x}'_g)
=
\frac{
\exp\!\big(\xi_g^{(n_{\mathrm{it}})}(\mathbf{x}'_g)\big)
}{
\sum\limits_{\mathbf{x}''_g \in \mathcal{X}_g}
\exp\!\big(\xi_g^{(n_{\mathrm{it}})}(\mathbf{x}''_g)\big)
}.
\end{equation}
Next, the group-wise local MAP estimate is given by
\begin{equation}\label{eq-42}
\hat{\mathbf{x}}'_g
=
\arg\max_{\mathbf{x}'_g \in \mathcal{X}_g}
\operatorname{Pr}_{\mathrm{MAP}}^{(n_{\mathrm{it}})}(\mathbf{x}'_g).
\end{equation}
In addition to the hard decision, soft information is required for iterative
message passing. Therefore, the estimate probability for $x\in\mathcal{S}$ can be obtained from the Walsh-constrained layer as
\begin{equation}\label{eq-43}
{\rm{Pr}}_{\mathrm{Walsh}}^{(n_{\mathrm{it}})}(z[c]=x)
=
\sum_{\substack{\mathbf{x}'_g \in \mathcal{X}_g, z_g[t]=x}}
\operatorname{Pr}_{\mathrm{MAP}}^{(n_{\mathrm{it}})}(\mathbf{x}'_g).
\end{equation}
To enhance numerical stability and avoid oscillatory behavior, a damping-based
update is applied by combining the above probability with the GMP output.
Specifically, the prior probability for the next iteration is updated as
\begin{align}\label{eq-44}
\operatorname{Pr}^{(n_{\mathrm{it}})}(z[c]=x)
=&
\eta\,
{\operatorname{Pr}}_{\mathrm{Walsh}}^{(n_{\mathrm{it}})}(z[c]=x)
\nonumber\\
&+
(1-\eta)\,
\operatorname{Pr}_{\mathrm{GMP}}^{(n_{\mathrm{it}})}(z[c]=x),
\end{align}
where ${\eta}\in{(0,1]}$ denotes the damping factor. The damping factor can achieve the trade-off between the convergence speed, stability and overall performance. Finally, 
the updated probability $\operatorname{Pr}^{(n_{\mathrm{it}})}(z[c]=x)$ is 
fed back to the first layer as the prior information for the $(n_{\mathrm{it}}+1)$-th
iteration.}

\begin{algorithm}[t]
\caption{Two-Layer Message Passing Detector with Walsh Constraints (TLMP-WC)}
\label{alg:tlmp-wc-short}
\begin{algorithmic}[1]
\Require $\mathbf{y}, \mathbf{H}_{\rm eff}^{\rm B}, \mathbf{W}_M, N_0, G, M, \mathcal{S}, n_{\rm it}^{\max}, \varepsilon, \eta$
\Ensure Estimated symbols $\widehat{\mathbf{x}}', \hat{\mathbf{z}}$

\State \textbf{Initialization:}
\State Set $\Pr^{(0)}(z[c]=x)=1/|\mathcal{S}|,\ \forall c,x$;
\State Initialize $n_{\rm it}=1$ and $\psi^{(0)}=0$;

\vspace{0.3em}
\While{$n_{\rm it}\le n_{\rm it}^{\max}$}

  \State \textbf{Gaussian Message Passing-Based Linear Detection:}
  \For{$c=1,\ldots,N$}
    \State Compute $\mu_{d,e}^{(n_{\rm it})}$ and $\left(\sigma_{d,e}^{(n_{\rm it})}
    \right)^2 \gets$~\eqref{eq-27} and~\eqref{eq-28};
    \State Evaluate ${\rm Pr}(y[d]\mid z[c]=x)\gets$~\eqref{eq-29};
    \State Compute $\Pr_{\rm GMP}^{(n_{\rm it})}(z[c]=x)\gets$~\eqref{eq-30};
    \State Update $\mu_c^{(n_{\rm it})}$ and $\left(\sigma_c^{(n_{\rm it})}\right)^2\gets$~\eqref{eq-32} and~\eqref{eq-33};
  \EndFor

  \vspace{0.3em}
  \State \textbf{Walsh-constrained Grouped Local MAP Detection:}
  \For{$g=1,\ldots,G$}
    \State Compute $\mu_{d,g}^{(n_{\rm it})}$ and $\left(\sigma_{d,g}^{(n_{\rm it})}\right)^2\gets$~\eqref{eq-35} and~\eqref{eq-36};
    \State Evaluate $\xi_g^{(n_{\rm it})}(\mathbf{x}'_g)\gets$~\eqref{eq-40};
    \State Compute $\Pr_{\rm MAP}^{(n_{\rm it})}(\mathbf{x}'_g)\gets$~\eqref{eq-41};
    \State Compute  ${\rm Pr}_{\rm Walsh}^{(n_{\rm it})}(z[c]=x)\gets$~\eqref{eq-43};
  \EndFor
  \State Update $\Pr^{(n_{\rm it})}(z[c]=x)\gets$~\eqref{eq-44};

  \vspace{0.3em}
  \State Compute
  $\psi^{(n_{\rm it})}\gets~\eqref{eq:conv_indicator}$;
  \If{$\psi^{(n_{\rm it})} \ge \psi^{(n_{\rm it}-1)}$}
    \State ${\rm Pr}(z[c]=x)\gets
{\rm Pr}^{(n_{\rm it})}(z[c]=x)$
  \EndIf
  \If{$\psi^{(n_{\rm it})}=1$ \textbf{or} $n_{\rm it}=n_{\rm it}^{\max}$}
    \State \textbf{break}
  \EndIf
  \State $n_{\rm it}\gets n_{\rm it}+1$
\EndWhile
\end{algorithmic}
\end{algorithm}
{\bf\em{3) Convergence indicator:}}
To characterize the convergence behavior of the proposed TLMP-WC detector, we
define the convergence indicator at the $n_{\rm it}$-th iteration as
\begin{align}\label{eq:conv_indicator}
\psi^{(n_{\rm it})}
=
\frac{1}{N}
\sum_{c=1}^{N}
\mathbb{I}
\left(
\max_{x\in\mathcal{S}}
{\rm Pr}^{(n_{\rm it})}(z[c]=x)
\ge 1-\varepsilon
\right),
\end{align}
where $\mathbb{I}(\cdot)$ denotes the indicator function and $\varepsilon>0$ is a
small predefined threshold.
The convergence indicator $\psi^{(n_{\rm it})}\in[0,1]$ represents the proportion
of symbol nodes whose posterior probabilities have sufficiently concentrated on a
single constellation point. A larger value of $\psi^{(n_{\rm it})}$ indicates a more stable detection~state.

\vspace{0.5em}
{\bf\em{4) Update criteria:}}
To prevent performance degradation caused by oscillatory message updates, the
symbol probabilities are updated only when the convergence indicator of the
current iteration is no worse than that of the previous one.
Specifically, the update criterion is given by
\begin{align}\label{eq:update_criterion}
{\rm Pr}(z[c]=x)
=
{\rm Pr}^{(n_{\rm it})}(z[c]=x),\!\!\!
\quad
\text{if }\;
\psi^{(n_{\rm it})}
\ge
\psi^{(n_{\rm it}-1)},
\end{align}
for all $c=1,\ldots,N$ and $x\in\mathcal{S}$.
Otherwise, the symbol probabilities remain unchanged.

{\bf\em{5) Stopping criteria and final decision:}}
The iterative TLMP-WC detector terminates when either the convergence indicator
reaches unity, i.e., $\psi^{(n_{\rm it})}=1$, or the maximum number of iterations
$n_{\rm it}^{\max}$ is reached.
After termination, the final detected symbol is obtained according to the
MAP criterion as
\begin{align}\label{eq:final_decision}
\hat{z}[c]
=
\arg\max_{x\in\mathcal{S}}
{\rm Pr}(z[c]=x),
\qquad c=1,\ldots,N.
\end{align}
The corresponding sparse symbol vector $\hat{\mathbf{x}}'$ is then recovered
based on the Walsh mapping relationship.
The pseudo-code of the proposed TLMP-WC detector is summarized in
{\bf Algorithm~\ref{alg:tlmp-wc-short}}.

\subsection{Computational Complexity Analysis}
We briefly analyze the computational complexity of the proposed TLMP-WC detector and compare it with the GMP~\cite{11185309}, and conventional MMSE-ML detectors~\cite{tao_JSAC}.
The computational complexity of the proposed TLMP-WC detector mainly
depends on the iterative message updates in the GMP-LD and WC-GLMD
layers. We quantify the dominant complexity in terms of real-valued
floating-point operations (FLOPs). Following the counting convention
in~\cite{tao_JSAC,Fan2018ILMP,Xiao2017ECS}, a complex multiplication,
a complex addition, and a complex-by-real multiplication are counted
as six, two, and two FLOPs, respectively, while each real
multiplication, addition, subtraction, or division is counted as one
FLOP. Exponential evaluations, comparisons, indexing operations,
final hard decisions, and one-time initialization operations are not
included.
Let $Q_{\rm s}=|\mathcal S|$,
$E=\operatorname{nnz}(\mathbf H_{\rm eff}^{\rm B})=Nd_c$, and
$B=\sum_{g=1}^{G}|\mathcal L_g|
=G\bar L=N\bar L/M$, where $\operatorname{nnz}(\cdot)$ denotes the number of nonzero entries of a matrix,
$d_c$ denotes the average column degree of
$\mathbf H_{\rm eff}^{\rm B}$ and $\bar L$ denotes the average number
of observation nodes associated with one Walsh group.
In each iteration, the GMP-LD layer computes the edge-excluded
interference statistics and symbol likelihoods according to
\eqref{eq-27}-\eqref{eq-33}. By reusing the row-wise statistics, its
computational complexity is calculated as
$F_{\mathrm{GMP\text{-}LD}}^{\rm iter}
=
E(9Q_{\rm s}+13)+N(14Q_{\rm s}-1)$. The WC-GLMD layer evaluates $MQ_{\rm s}$ valid symbol-index candidates
for each Walsh group and updates their probabilities according to
\eqref{eq-35}-\eqref{eq-44}. By reusing the group-wise statistics,
its computational complexity can be given by
$F_{\mathrm{WC\text{-}GLMD}}^{\rm iter}
={}
10E+B(7MQ_{\rm s}+5)+NQ_{\rm s}(M+4)-N-G$.
Therefore, with $n_{\rm it}^{T}$ denoting the number of TLMP-WC iterations,
the computational complexity is
$F_{\rm TLMP-WC}
=\left(F_{\mathrm{GMP\text{-}LD}}^{\rm iter}
+F_{\mathrm{WC\text{-}GLMD}}^{\rm iter}\right)n_{\rm it}^{T}$.

For comparison, let $\widetilde d_c
=\operatorname{nnz}(\mathbf H_{\rm eff}^{\rm B}\mathbf W_{\rm eff})/N$
denote the average column degree of the composite channel employed by
GMP. Following the single-layer MP count in~\cite{tao_JSAC}, its
computational complexity is
$F_{\rm GMP}
=\left[N\widetilde d_c(31Q_{\rm s}+43)-2N\right]n_{\rm it}^{G}$,
where $n_{\rm it}^{G}$ denotes the number of GMP iterations.
For the MMSE-ML detector, the MMSE equalization requires
$16N^3+13N^2$ FLOPs under the same counting convention
in~\cite{tao_JSAC}. Including the subsequent group-wise ML search
in~\cite{tao2023affine}, its computational complexity is
$F_{\rm MMSE-ML}
=16N^3+13N^2+GKQ_{\rm s}(12M-1)$.
We can find that for fixed graph degrees, group size, constellation size, and 
iteration numbers, TLMP-WC and GMP scale linearly with $N$, whereas
MMSE-ML is dominated by the cubic MMSE operation. Numerical FLOP
comparisons and the corresponding tradeoff between the BER performance and complexity are
presented in Section~VI.

\begin{table}[t]
\centering
\caption{Simulation Parameters}
\label{tab:simulation_parameters}
\centering
\renewcommand\arraystretch{1.2}
\setlength{\tabcolsep}{6mm}
\begin{tabular}{|c|c|}
\hline
\textbf{Parameter} & \textbf{Value} \\
\hline
Carrier frequency $f_c$ & $10~\mathrm{GHz}$ \\ \hline
Chirp subcarrier spacing $\Delta f$ & $1.5~\mathrm{kHz}$ \\ \hline
Number of chirp subcarriers $N$ & $64$ \\ \hline
Number of subcarriers per group $M$ & $4$ \\ \hline
Modulation scheme & QPSK \\ \hline
Post-chirp parameter $c_1$ & $\frac{2(\alpha_{\rm max}+1)+1}{2N}$ \\ \hline
Maximum normalized delay $l_{\max}$ & $2$ \\ \hline
Number of channel taps $P$ & $3$ \\ \hline
Maximum normalized Doppler $\alpha_{\max}$ & $3$ \\ \hline
Maximum mobile speed $v_{\max}$ & $486~\mathrm{km/h}$ \\ \hline
Dead-zone parameter $k$ & $1.5$ \\ \hline
Step-size parameter $\gamma$ & $2$ \\ \hline
Scaling factor $\alpha$ & $0.1$ \\ \hline
Non-zero quantization levels $L$ & $4$ \\ \hline
\end{tabular}
\end{table}
\section{Simulation Results and Discussions}
In this section, we evaluate the reliability and security performance of the proposed AFDM-PCS-IM scheme, as well as the effectiveness of the proposed low-complexity TLMP-WC detector over LTV channels. For clarity, the simulation parameters are summarized in Table~\ref{tab:simulation_parameters}.
Without loss of generality, the carrier frequency is set to $10~\mathrm{GHz}$ and the chirp subcarrier spacing is chosen as
$\Delta f=1.5~\mathrm{kHz}$. Unless otherwise specified, the maximum normalized integer Doppler shift is set to ${\alpha}_{\rm max}=3$, which corresponds to a maximum mobile speed of approximately $486~\mathrm{km/h}$~\cite{11185309}. 
The maximum normalized delay is ${l}_{\rm max}=2$ and the number of channel taps is $P=3$. Besides, the total number of chirp subcarriers is set to $N = 64$, the number of chirp subcarriers per group is set to $M = 4$, and QPSK mapping is employed.  
For the proposed CSI quantizer, the dead-zone width and the step size are set
to $k=1.5$ and $\gamma=2$, yielding $d=1.5\sigma_h$ and
$\Delta=2\sigma_h$, respectively. This setting provides a favorable
tradeoff between encrypted codebook consistency and CSI quantization
resolution. Specifically, the dead zone suppresses noise-dominated
low-magnitude CSI components, while the selected step size improves
tolerance to estimation mismatch without excessively coarsening the
quantized CSI representation. The scaling factor is set to $\alpha=0.1$, and the number of non-zero quantization levels is chosen as~$L = 4$.

\subsection{Effectiveness of The Proposed Codebook and Quantizer Design}
\begin{figure}[t]
	\center
	\includegraphics[width=2.5in,height=1.8in]{{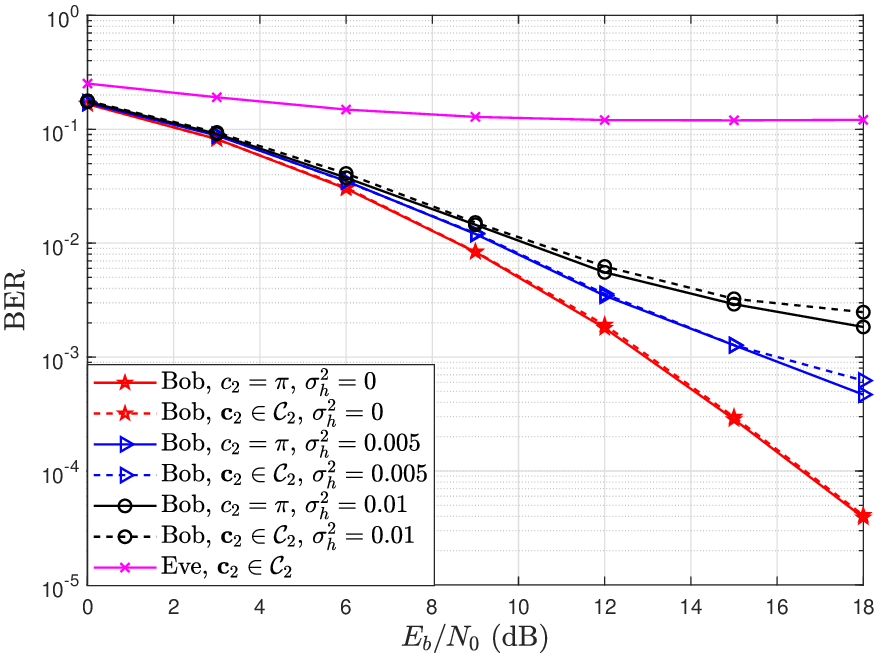}}
	\vspace{-0.0cm}
	\caption{{BER performance of the proposed AFDM-PCS-IM scheme under different pre-chirp parameter settings and CSI estimation error levels.}}
	\label{effect-coodedbook}  
	\vspace{-1mm}
\end{figure}

In this subsection, an MMSE-ML~\cite{tao_JSAC} detection algorithm is employed for demodulation.
It should be noted that the ``covertness'' refers to preventing Eve from
reconstructing the legitimate CSI-encrypted pre-chirp codebook and
recovering the information embedded in it. Accordingly, it is evaluated
jointly through the pre-chirp codebook match rate and Eve's BER, where a
low codebook match rate together with a persistently high BER floor
indicates unsuccessful codebook reconstruction and information
recovery at Eve.
First, we evaluate the BER performance of the proposed AFDM-PCS-IM scheme under different pre-chirp parameter settings and CSI estimation error levels. We consider two configurations for the pre-chirp parameter: (i) a fixed pre-chirp parameter with $c_{2} = \pi$, and (ii) the proposed pre-chirp-parameter-based encryption strategy, where $\mathbf{c}_{2} \in \mathcal{C}_{2}$.
As shown in Fig.~\ref{effect-coodedbook}, the proposed AFDM-PCS-IM with $\mathbf{c}_{2} \in \mathcal{C}_{2}$ achieves a BER performance close to that of AFDM-PCS-IM with $c_{2} = \pi$ across all CSI error levels. In addition, the BER performance of both schemes gradually degrades as the CSI estimation error increases. This indicates that, although the proposed scheme relies heavily on accurate CSI, it still maintains good robustness in the presence of CSI estimation errors. Meanwhile, since Eve cannot acquire the pre-chirp parameter codebook shared between Alice and Bob, her demodulation is almost equivalent to random guessing. Hence, Eve's BER maintains a persistently high error floor over the
considered SNR range, which is consistent with the design insights
derived in Section~\ref{codebook}.

\begin{figure}[t]
	\center
	\includegraphics[width=2.5in,height=1.8in]{{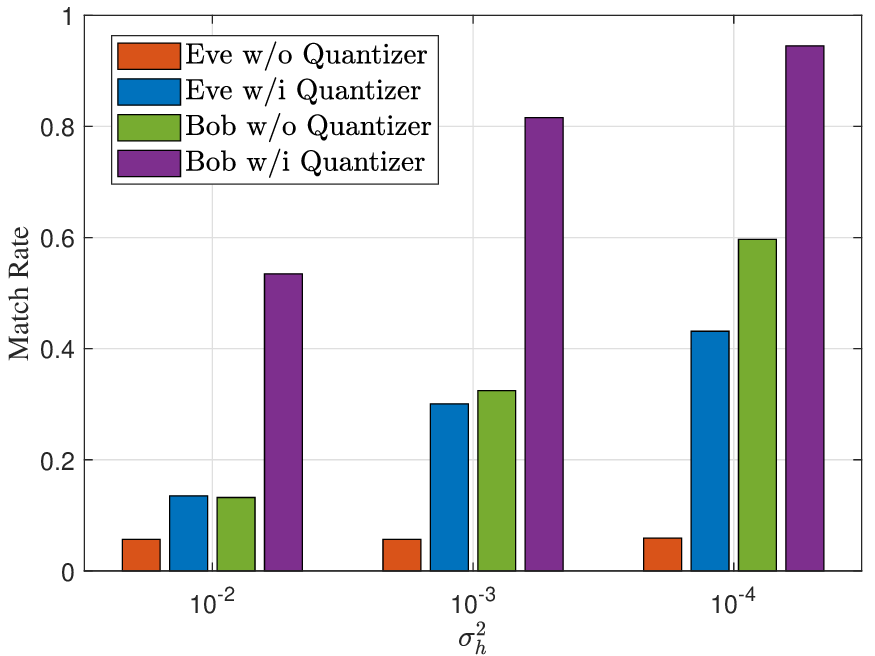}}
	\vspace{-0.0cm}
	\caption{{Pre-chirp parameters match rates of the proposed AFDM-PCS-IM scheme with/without quantizer under different CSI estimation error levels.}}
	\label{match-rate}  
	\vspace{-2mm}
\end{figure}
\begin{figure}[t]
	\center
	\includegraphics[width=2.5in,height=1.8in]{{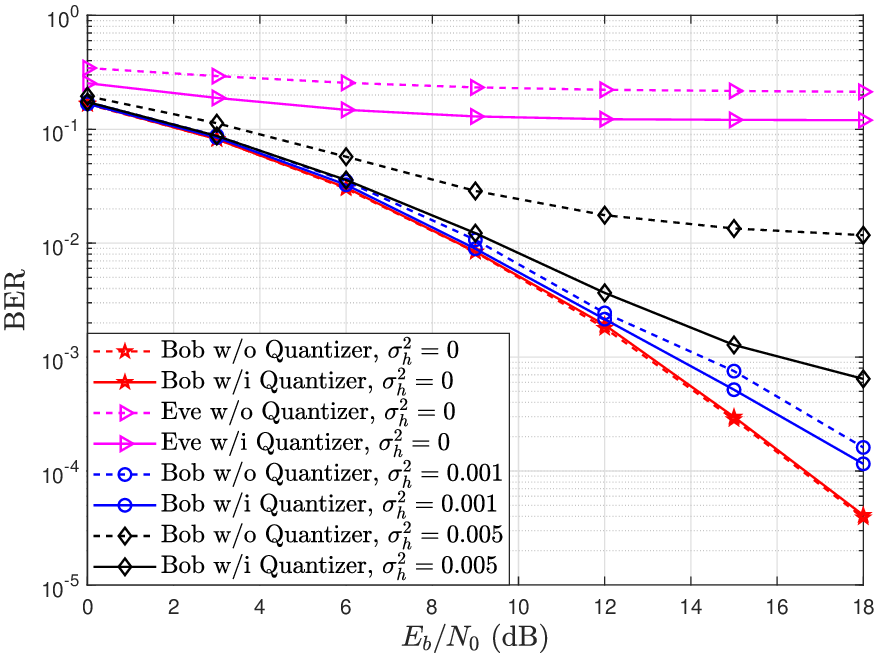}}
	\vspace{-0.0cm}
	\caption{{BER performance of the proposed AFDM-PCS-IM scheme with/without quantizer under different CSI estimation error levels.}}
	\label{effect-quantizer}  
	\vspace{-2mm}
\end{figure}
Next, we evaluate the effectiveness of the proposed noise-aware CSI quantizer. 
To evaluate the security performance
under a conservative threat model, we assume that Eve has full
knowledge of the Walsh-Hadamard codebook, the pre-chirp parameter
mapping, the CSI quantizer structure, the CSI-driven
codebook construction rule, and the receiver algorithm. Fig.~\ref{match-rate} illustrates the match rate of the pre-chirp parameters between Alice and the receivers under different CSI estimation error levels ${\sigma}_h^2$.
Here, the match rate measures the entry-wise agreement between the CSI-driven pre-chirp perturbation sequence generated by Alice and its reconstruction at the receiver. It is calculated as the fraction of matched entries averaged over all Monte Carlo frames.
A high match rate indicates that the receiver can
reliably reconstruct the legitimate CSI-encrypted pre-chirp codebook,
whereas a low match rate indicates a severe codebook mismatch.
For the legitimate user Bob, the proposed quantizer significantly improves the robustness of the AFDM-PCS-IM scheme against imperfect CSI. As the CSI estimation error increases, the match rate without quantization drops rapidly. In contrast, when the noise-aware CSI quantizer is applied, the degradation occurs much more slowly. 
For instance, when ${\sigma}_h^2 = 10^{-4}$, the match rate between Bob and Alice improves from approximately 0.6 to 0.94 with the proposed quantizer, yielding a gain of about 57$\%$.
This result indicates that the proposed noise-aware CSI quantizer effectively mitigates the impact of CSI estimation errors, allowing the legitimate link to maintain a high level of parameter consistency even under imperfect CSI conditions.

In contrast, for the eavesdropper Eve, the situation is totally different.
On the one hand, the quantizer involves multiple tunable parameters, making it difficult for Eve to obtain exactly the same configuration as the legitimate pair. On the other hand, since Eve's channel is statistically independent of the Alice $\to$ Bob link, even if part of the quantizer settings were exposed, her matching probability with Alice would still remain low. Therefore, while the proposed noise-aware CSI quantization mechanism substantially enhances the robustness of the legitimate link, it does not provide a comparable benefit to the~eavesdropper.

 We further investigate the impact of the proposed noise-aware CSI quantizer on the BER performance of the AFDM-PCS-IM system. In Fig.~\ref{effect-quantizer}, we give the BER results of the proposed AFDM-PCS-IM scheme with and without the quantizer under different CSI estimation error levels. When perfect CSI is available (i.e., ${\sigma}_h^2 = 0$), the BER curves with and without a quantizer almost coincide, indicating that the proposed quantizer does not introduce any performance loss in the ideal case. 
However, under imperfect CSI conditions, the advantage of the quantizer becomes more pronounced. For ${\sigma}_h^2 = 0.001$ and ${\sigma}_h^2 = 0.005$, the BER without quantization degrades significantly and exhibits an error floor at high SNR, due to the pre-chirp parameters mismatch caused by CSI estimation errors. In contrast, with the proposed quantizer, the BER performance is substantially improved, and the error floor is effectively suppressed. This behavior is consistent with the improved pre-chirp parameters match rate observed in Fig.~\ref{match-rate}. 
Importantly, Fig. 5 shows that applying the quantizer does not bring any noticeable BER improvement at Eve.
This indicates that the proposed noise-aware CSI quantizer primarily benefits the legitimate receiver without significantly altering the eavesdropper's detection conditions. 

\begin{figure}[t]
	\center
	\includegraphics[width=2.5in,height=1.8in]{{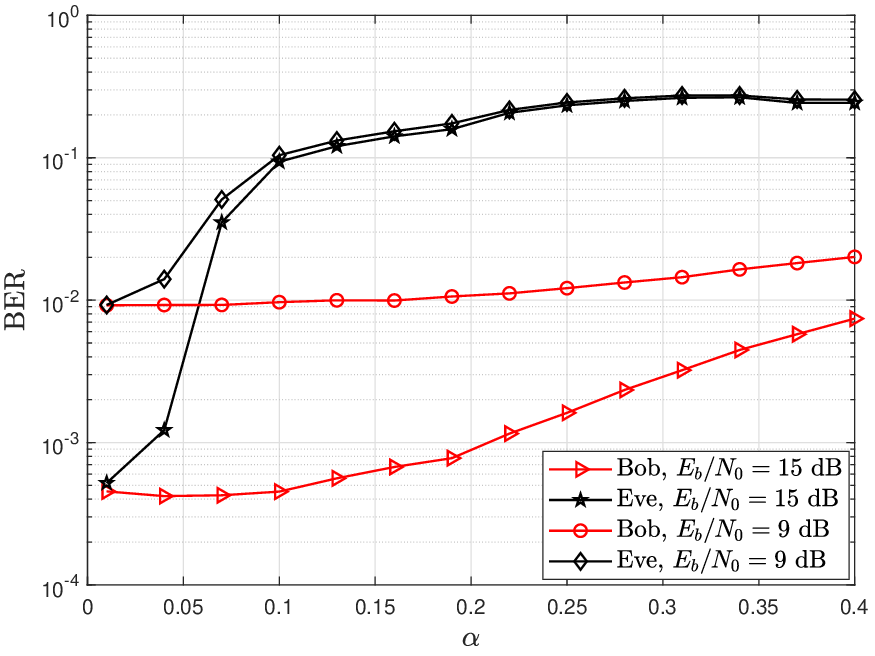}}
	\vspace{-0.0cm}
	\caption{{BER performance of the proposed AFDM-PCS-IM scheme versus the scaling factor $\alpha$ at Bob and Eve, where $\sigma^2_h=0.001$.}}
	\label{effect-alpha}  
	\vspace{-1mm}
\end{figure}
Fig.~\ref{effect-alpha} illustrates the impact of the scaling factor $\alpha$ on the BER performance. According to~\eqref{q-19}-\eqref{eq:CSI-quantizer}, $\alpha$
determines the spacing between adjacent nonzero output labels $q_i$ and
thereby controls the strength of the CSI-driven perturbation.
As $\alpha$ increases, the separation between adjacent nonzero output
labels becomes larger, thereby strengthening the CSI-driven
perturbation.
As shown in Fig.~\ref{effect-alpha}, when $\alpha$ is small (e.g., $\alpha < 0.05$), the spacing between adjacent nonzero output
labels remains limited and the resulting pre-chirp parameter mismatch is relatively mild. In this case, Eve's BER increases only slightly, while Bob maintains a low BER. This indicates that the reliability of the legitimate link is preserved, but the security enhancement is limited.
As $\alpha$ increases, Eve's BER rises noticeably, since the larger separation between adjacent nonzero output labels leads to more severe pre-chirp parameter mismatch at the eavesdropper and hence poorer detection performance. However, when $\alpha$ becomes excessively large (e.g., $\alpha > 0.2$), Bob's BER also starts to increase. Under imperfect CSI, a large $\alpha$ enlarges the pre-chirp parameter mismatch even for the legitimate link, which degrades Bob's pre-chirp codebook reconstruction and detection performance.
Overall, the choice of $\alpha$ reflects a trade-off between reliability and security. A too small $\alpha$ fails to introduce sufficient perturbation, whereas a too large $\alpha$ impairs the legitimate receiver. As observed in Fig.~\ref{effect-alpha}, around $\alpha = 0.1$, Bob's BER remains low. Meanwhile, Eve's BER is already significantly elevated, providing a favorable balance. Therefore, we employ $\alpha = 0.1$ in the simulations to achieve an expected trade-off between reliability and security.

\begin{figure}[t]
	\center
	\includegraphics[width=2.5in,height=1.8in]{{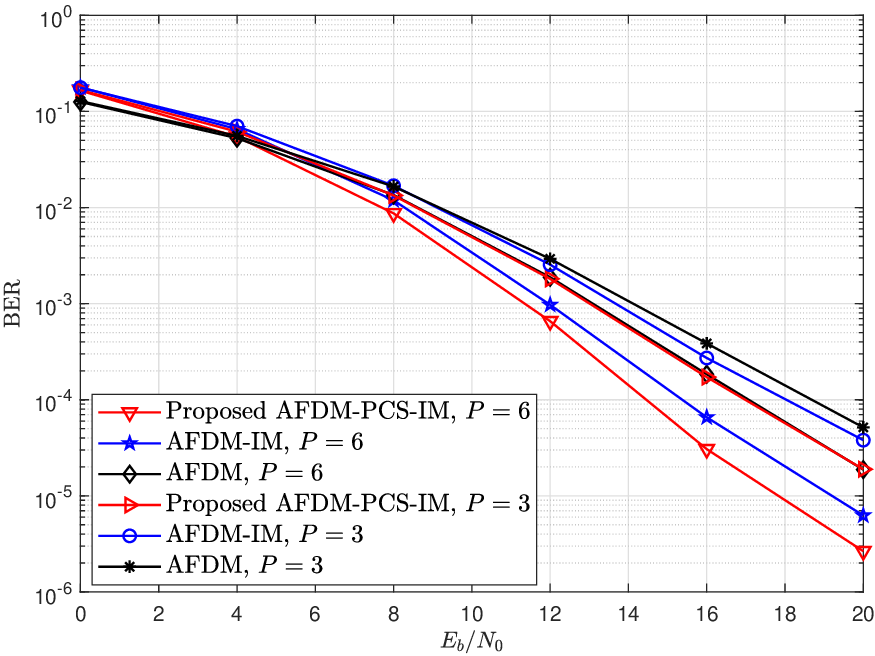}}
	\vspace{-0.0cm}
	\caption{{BER performance comparison of the proposed AFDM-PCS-IM, AFDM, and AFDM-IM schemes with perfect CSI at Bob.}}
	\label{com-scheme}  
	\vspace{-1mm}
\end{figure}

In Fig~\ref{com-scheme}, we provide the BER performance at Bob and Eve for the
proposed AFDM-PCS-IM scheme, in comparison with the conventional AFDM~\cite{10087310}
and AFDM-IM~\cite{tao_JSAC} schemes under perfect CSI. To ensure a fair comparison, we fix the data rate at 1 $\rm bps/Hz$. Thus, the proposed AFDM-PCS-IM scheme and AFDM-IM scheme use the QPSK mapping, while AFDM scheme uses the BPSK mapping. As shown in Fig.~\ref{com-scheme},  the
proposed AFDM-PCS-IM scheme achieves superior BER
performance at Bob than both AFDM and AFDM-IM schemes. For instance, when the target BER is $10^{-4}$ and $P=6$, the
proposed AFDM-PCS-IM scheme achieves about 1 $\rm dB$ and 2.5 $\rm dB$ performance
gains over the AFDM-IM and AFDM schemes, respectively.
This further demonstrates the high reliability of the proposed AFDM-PCM-IM scheme and the effectiveness of the codebook design in Section~\ref{codebook}.

\begin{figure}[t]
	\center
	\includegraphics[width=2.5in,height=1.8in]{{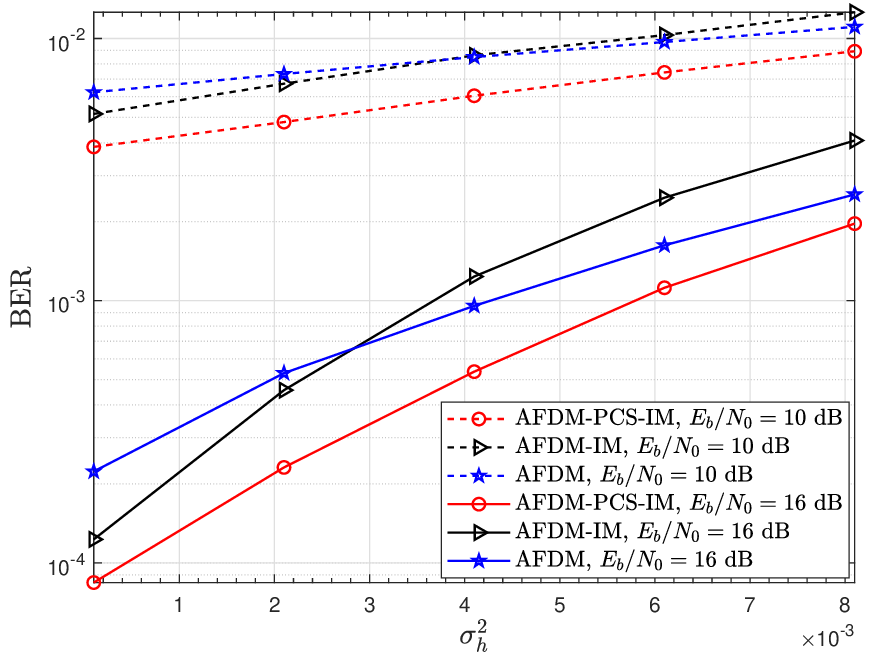}}
	\vspace{-0.0cm}
	\caption{{BER performance versus the channel estimation errors for the proposed AFDM-PCS-IM
    scheme and the benchmark AFDM and AFDM-IM schemes at Bob.}}
	\label{sim3}  
	\vspace{-0mm}
\end{figure}

In Fig.~\ref{sim3}, we compare the proposed AFDM-PCS-IM scheme with the conventional AFDM scheme and the AFDM-IM scheme under channel estimation errors at Bob. It can be observed that the proposed AFDM-PCS-IM scheme achieves superior BER performance over both benchmark schemes across the low-SNR and medium-to-high-SNR regions.
When the channel estimation error variance $\sigma_h^2$ is small, e.g., around $10^{-3}$, the AFDM-IM scheme outperforms the conventional AFDM scheme due to the additional index modulation gain. Under the same condition, the proposed AFDM-PCS-IM scheme further reduces the BER by approximately one order of magnitude compared with the conventional AFDM scheme.

Moreover, we can observe that as $\sigma_h^2$ increases, the BER of all three schemes degrades, but with different sensitivities to CSI mismatch. In particular, when $\sigma_h^2$ exceeds approximately $4\times 10^{-3}$, the BER of the AFDM-IM scheme increases more rapidly and eventually becomes worse than that of the conventional AFDM scheme. This phenomenon is attributed to the detection structure of the AFDM-IM scheme. Since information is jointly conveyed by index bits and modulation bits, reliable detection relies heavily on accurate CSI. Channel estimation errors make the identification of active resource elements less reliable, resulting in incorrect index decisions and consequently deteriorating the overall BER performance.
In contrast, although the BER of the proposed AFDM-PCS-IM scheme also increases with $\sigma_h^2$, its BER remains consistently lower than those of both the conventional AFDM scheme and the AFDM-IM scheme across the entire error range. This indicates that the proposed design is more resilient to imperfect CSI. Even under relatively severe channel estimation errors, the proposed AFDM-PCS-IM scheme maintains a clear performance advantage, thereby demonstrating its robustness in practical imperfect CSI scenarios.

\begin{figure}[t]
	\center
	\includegraphics[width=2.5in,height=1.8in]{{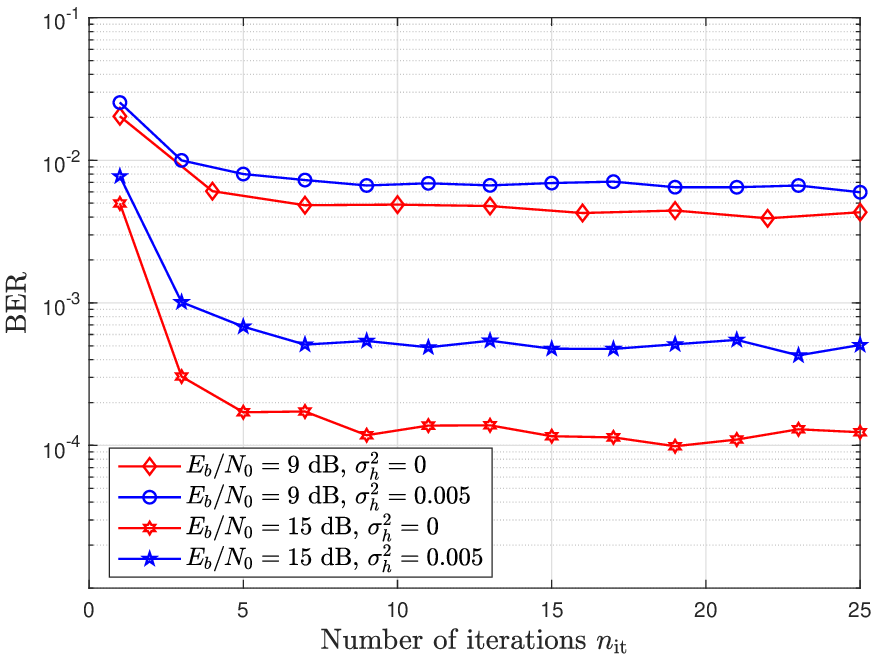}}
	\vspace{-0.0cm}
	\caption{{BER performance of the proposed AFDM-PCS-IM scheme with TLMP-WC detector versus the number of iterations ${n}_{\rm it}$.}}
	\label{num-iter}  
	\vspace{-0mm}
\end{figure}
\subsection{Effectiveness of the Proposed TLMP-WC Detector}
Fig.~\ref{num-iter} illustrates the BER performance of the proposed AFDM-PCS-IM scheme with the TLMP-WC detector versus the number of iterations $n_{\mathrm{it}}$ under different $E_b/N_0$ and CSI error conditions.
It can be observed that the BER decreases rapidly during the first several iterations and gradually converges as $n_{\mathrm{it}}$ increases. In particular, most of the performance improvement is achieved within the first $10-15$ iterations. Beyond this range, further increasing $n_{\mathrm{it}}$ yields only marginal gains, and the BER curves become nearly flat.
These results verify the convergence behavior of the proposed TLMP-WC detector. The iterative message-passing process effectively refines the symbol estimates in the early iterations, while later iterations mainly provide minor performance improvements. 
Based on this observation, $n_{\mathrm{it}} = 15$ is adopted in subsequent simulations, providing a favorable trade-off between detection performance and computational complexity. 
\begin{figure}[t]
	\center
	\includegraphics[width=2.5in,height=1.8in]{{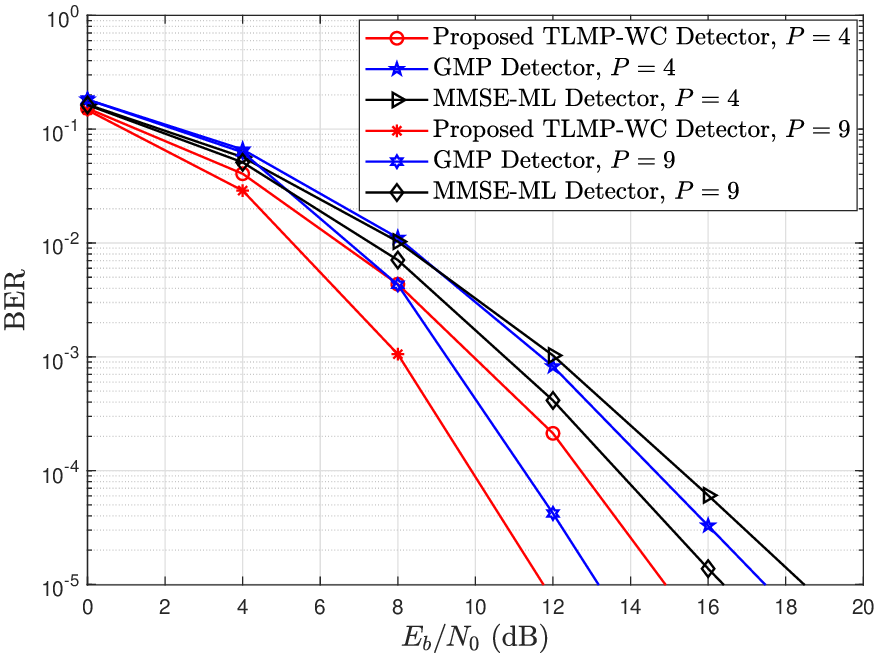}}
	\vspace{-0.0cm}
	\caption{{BER performance of the proposed TLMP-WC, GMP and MMSE-ML detectors for the AFDM-PCS-IM scheme with perfect CSI at Bob.}}
	\label{com-detector}  
	\vspace{-0mm}
\end{figure}
Fig.~\ref{com-detector} compares the BER performance of the proposed AFDM-PCS-IM scheme under different detection algorithms, including the proposed TLMP-WC detector, the GMP detector, and the MMSE-ML detector, assuming perfect CSI at Bob.
The proposed TLMP-WC detector consistently outperforms the GMP and MMSE-ML detectors across the entire SNR range. For example, at a target BER of $10^{-3}$ and $P=4$, the proposed TLMP-WC detector achieves approximately $1.8$~dB and $2$~dB gains compared with the GMP and MMSE-ML detectors, respectively. Similar trends are observed for $P=9$, where TLMP-WC provides about $1.5$~dB performance gain over GMP at the same BER level. 
Moreover, the performance gap becomes more evident in the medium-to-high SNR region, where the BER curves of TLMP-WC exhibit a steeper decay. This demonstrates that the proposed two-layer message passing algorithm more effectively exploits the sparsity structure and the pre-chirp codebook characteristics of the AFDM-PCS-IM~scheme. 
\begin{table}[t]
\centering
\caption{Computational complexity comparison
among different detectors, where $N=64$, $M=4$, and QPSK modulation is
employed.}
\label{tab:numerical-flops}
\renewcommand{\arraystretch}{1.15}
\setlength{\tabcolsep}{9pt}
\begin{tabular}{lcc}
\toprule
& \multicolumn{2}{c}{FLOPs per detected AFDM block}\\
\cmidrule(lr){2-3}
Detector & $P=4$ & $P=9$\\
\midrule
TLMP-WC & $6.85\times10^{5}$ & $1.22\times10^{6}$\\
GMP     & $2.15\times10^{6}$ & $3.59\times10^{6}$\\
MMSE-ML & $4.26\times10^{6}$ & $4.26\times10^{6}$\\
\bottomrule
\end{tabular}
\end{table}

To quantitatively evaluate the tradeoff between the BER performance and complexity,
Table~\ref{tab:numerical-flops} reports the receiver-side FLOP counts
of the considered detectors under the simulation parameters used in
Fig.~\ref{com-detector}. The graph parameters are averaged over 1000 independent channel
realizations. Specifically, the resulting
$(d_c,\bar L,\widetilde d_c)$ values are
$(4,13.401,13.401)$ for $P=4$ and
$(9,22.405,22.405)$ for $P=9$. Based on the convergence results in Fig.~\ref{num-iter}, a common fixed number of iterations, i.e., $n_{\rm it}^{T}=n_{\rm it}^{G}=15$ is adopted for TLMP-WC and GMP.
For $P=4$, TLMP-WC requires approximately
$6.85\times10^{5}$ FLOPs per detected AFDM block, corresponding to
complexity reductions of $68.1\%$ and $83.9\%$ relative to GMP and
MMSE-ML, respectively. For $P=9$, the corresponding reductions are
approximately $66.0\%$ and $71.3\%$, respectively. This reduction
mainly stems from the fact that TLMP-WC processes the sparse channel
matrix $\mathbf{H}_{\rm eff}^{\rm B}$ directly, whereas GMP operates
on the less sparse composite matrix
$\mathbf{H}_{\rm eff}^{\rm B}\mathbf{W}_{\rm eff}$. Together with the
BER gains observed in Fig.~\ref{com-detector}, these results demonstrate
that TLMP-WC achieves a favorable tradeoff between detection
performance and computational complexity under the considered
simulation~settings.

\begin{figure}[t]
	\center
	\includegraphics[width=2.5in,height=1.8in]{{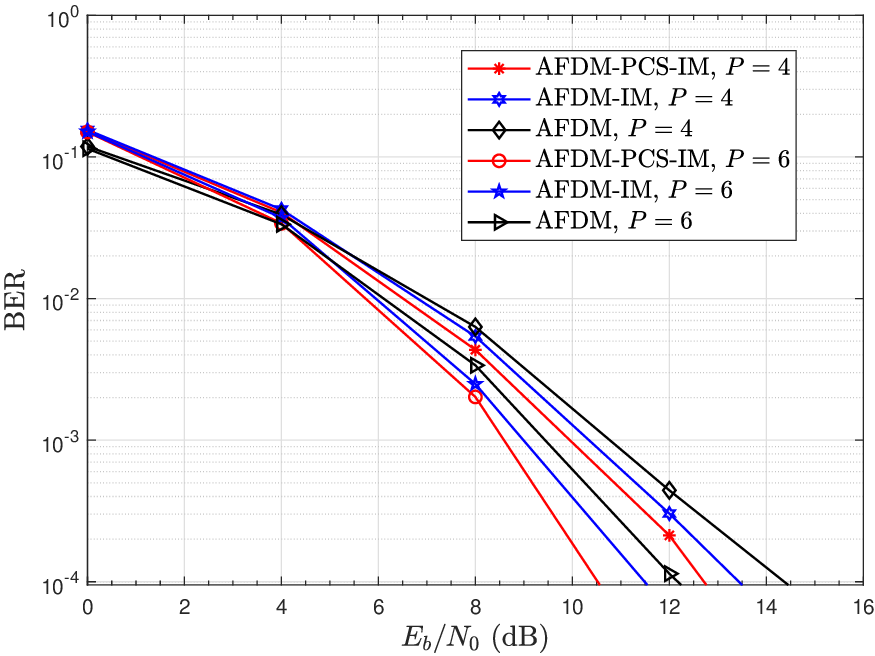}}
	\vspace{-0.0cm}
	\caption{{{BER performance comparison of the proposed AFDM-PCS-IM, AFDM-IM, and conventional AFDM schemes with the advanced MP detectors.}}}
	\label{com-scheme-MP}  
	\vspace{-0mm}
\end{figure}

In Fig.~\ref{com-scheme-MP}, we illustrate the BER performance of the AFDM-PCS-IM, AFDM-IM, and conventional AFDM schemes. The AFDM-PCS-IM scheme employs the proposed TLMP-WC detector, the AFDM-IM scheme adopts the DLMP detector~\cite{tao_JSAC}, and the conventional AFDM scheme is implemented with the GMP detector~\cite{11185309}.
It is worth emphasizing that all three detectors are derived from the Gaussian message passing framework and operate with similar computational complexity orders, thereby ensuring a fair comparison.
As observed from the figure, the proposed AFDM-PCS-IM scheme achieves the lowest BER among the three schemes for both channel conditions.
For instance, at a target BER of $10^{-4}$ and $P=4$, the proposed AFDM-PCS-IM provides approximately 0.8 dB gain compared with AFDM-IM and about 1.8 dB gain compared with conventional AFDM. A similar trend is observed for $P=6$, where the proposed AFDM-PCS-IM scheme maintains an SNR advantage of roughly $1\sim 2$ dB at the same BER level.
Moreover, increasing the number of channel paths from $P=4$ to $P=6$ slightly improves the BER performance of all schemes due to enhanced multipath diversity. Nevertheless, the relative performance ordering remains unchanged, demonstrating the robustness of the proposed AFDM-PCS-IM scheme under the doubly dispersive channel conditions.

\begin{figure}[t]
	\center
	\includegraphics[width=2.5in,height=1.8in]{{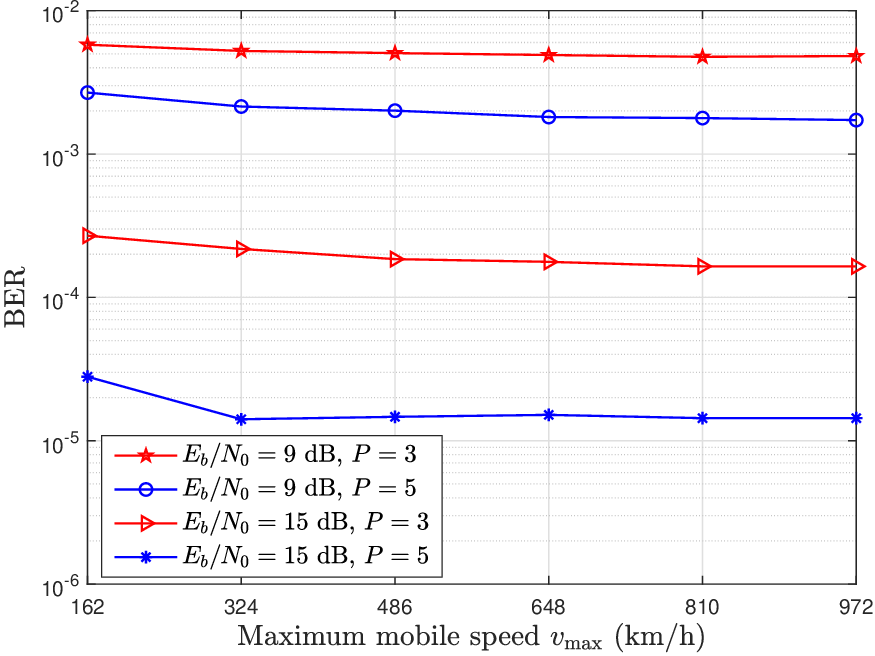}}
	\vspace{-0.0cm}
	\caption{{BER performance versus the maximum mobile speeds for the proposed AFDM-PCS-IM scheme at Bob.}}
	\label{speed}  
	\vspace{-0mm}
\end{figure}

Finally, we further evaluate the performance of the proposed AFDM-PCS-IM scheme in different velocity scenarios. Fig.~\ref{speed} presents the BER performance at Bob under different maximum mobile speeds $v_{\max}$, where $v_{\max}$ increases from 162~km/h to 972~km/h. This speed range covers typical high-mobility communication scenarios, such as high-speed railway systems, vehicular networks, and low-altitude UAV communications.
As observed from Fig.~\ref{speed}, when the mobile speed increases from 162~km/h to 486~km/h, a slight BER improvement can be observed. This phenomenon can be explained from the perspective of the Doppler representation structure in the DAFT domain. 
At relatively low mobile speeds, the maximum normalized Doppler range is small, which results in a reduced spacing among different channel taps on the DAFT domain grid. In this case, fractional Doppler dispersion becomes more pronounced. Consequently, fractional Doppler multipath components from different channel taps overlap in the DAFT domain grids, reducing the delay-Doppler resolvability.
As the mobile speed increases, the maximum Doppler shift enlarges while the number of channel paths remains unchanged.
Therefore, the effective spacing among different channel taps in the DAFT domain increases. The path-overlapping effect caused by fractional Doppler dispersion is alleviated, thereby improving channel resolvability and enabling more effective exploitation of diversity gain.
Moreover, when the mobile speed further increases beyond 486~km/h, the BER curves tend to become stable. 
Although the maximum Doppler shift continues to increase, the grid spacing between adjacent normalized Doppler taps remains sufficient to maintain a stable sparse structure of the channel in the DAFT domain.
Therefore, the proposed TLMP-WC detector operates under nearly identical sparsity conditions, and no additional performance improvement is observed.
Overall, these results show that the proposed AFDM-PCS-IM scheme can effectively exploit the delay-Doppler characteristics of high-mobility channels and maintains robust detection performance over a wide range of mobile speeds.

\vspace{-2mm}
\section{Conclusion}
In this paper, we proposed a covert pre-chirp spreading index modulation framework for AFDM systems to enable physical-layer secure communication in high-mobility scenarios. 
By exploiting the flexible configurability of AFDM pre-chirp parameters, we constructed an orthogonal pre-chirp parameter codebook and introduced a CSI-driven perturbation mechanism that tightly couples the modulation structure with instantaneous legitimate channel characteristics. 
Through this design, the encrypted pre-chirp codebook serves as an additional index modulation dimension, allowing implicit data encryption without requiring explicit secret key exchange between the transmitter and the legitimate receiver.
To enhance practical applicability, a noise-aware CSI quantization strategy was further developed to mitigate the impact of channel estimation errors, thereby ensuring reliable encrypted codebook reconstruction under imperfect CSI conditions. 
In addition, a low-complexity TLMP-WC detector was developed to efficiently support large-scale data detection while maintaining high reliability.
Finally, simulation results under high-mobility and imperfect CSI environments demonstrated that the proposed AFDM-PCS-IM framework achieves superior covertness, improved BER performance, and enhanced robustness compared with conventional AFDM and AFDM-IM schemes. 
The simulation results further reveal that the overall design tradeoff
can be divided into two aspects. The CSI quantization and perturbation
parameters determine the reliability-security balance by maintaining
reliable codebook reconstruction at Bob while preventing Eve from
reconstructing the legitimate codebook. Meanwhile, the receiver design
determines the BER-complexity tradeoff, where the proposed TLMP-WC
detector achieves improved BER performance with fewer FLOPs than the
considered benchmark detectors. Therefore, their joint configuration 
provides a favorable balance among reliability, security, and detection
complexity.
These findings confirm that leveraging the pre-chirp parameter domain as a CSI-adaptive modulation dimension offers a promising approach for unifying waveform design and physical-layer security in time-varying wireless environments.
In future work, we will investigate the extension of the proposed
framework to multi-user scenarios, including user-specific codebook
design and multi-user detection.

\vspace{-2mm}
\bibliographystyle{IEEEtran}  
\end{document}